\documentclass[twocolumn]{aastex631}

\usepackage{graphicx}	
\usepackage{amsmath}	
\usepackage{amssymb}	
\usepackage{longtable}
\usepackage{grffile}
\usepackage{ulem}
\usepackage{mathrsfs}
\usepackage{mathptmx}
\usepackage{xspace}
\usepackage{times}
\usepackage{enumitem}
\usepackage{soul}

\newcommand{\pc}{\ensuremath{{\rm pc}}\xspace}
\newcommand{\kpc}{\ensuremath{{\rm kpc}}\xspace}
\newcommand{\Mpc}{\ensuremath{{\rm Mpc}}\xspace}

\newcommand{\MHz}{\ensuremath{{\rm MHz}}\xspace}

\newcommand{\GHz}{\ensuremath{{\rm GHz}}\xspace}
\newcommand{\Jy}{\ensuremath{{\rm Jy}}\xspace}

\newcommand{\K}{\ensuremath{\,{\rm K}\ }}

\begin{document}

\title{A Multi-Mechanism Hybrid Model of Peaked-Spectrum Radio Sources} 

\author[0000-0002-1398-5588]{Guang-Chen Sun}
\affiliation{Key Laboratory of Cosmology and Astrophysics (Liaoning)\\
\& College of Sciences, Northeastern University, Shenyang 110819, China\\}
\affiliation{National Astronomical Observatories, Chinese Academy of Sciences, Beijing 100101, China\\}
\affiliation{School of Astronomy and Space Science, University of Chinese Academy of Sciences, Beijing 100049, China\\}

\correspondingauthor{Yichao Li}
\author[0000-0003-1962-2013]{Yichao Li}
\email{liyichao@mail.neu.edu.cn}
\affiliation{Key Laboratory of Cosmology and Astrophysics (Liaoning)\\
\& College of Sciences, Northeastern University, Shenyang 110819, China\\}

\author[0000-0001-8075-0909]{Furen Deng}
\affiliation{National Astronomical Observatories, Chinese Academy of Sciences, Beijing 100101, China\\}
\affiliation{School of Astronomy and Space Science, University of Chinese Academy of Sciences, Beijing 100049, China\\}
\affiliation{Key Laboratory of Radio Astronomy and Technology, Chinese Academy of Sciences, A20 Datun Road, Chaoyang District, Beijing 100101, China}
\affiliation{Institute of Astronomy, University of Cambridge, Madingley Road, Cambridge, CB3 0HA, UK\\}

\author[0000-0002-4456-6458]{Yanping Cong}
\affiliation{Shanghai Astronomical Observatory, Chinese Academy of Sciences, Shanghai 200030, China\\}

\author[0000-0001-7943-0166]{Fangxia An}
\affiliation{Purple Mountain Observatory, Chinese Academy of Sciences, 10 Yuanhua Road, Qixia District, Nanjing 210023, China\\}
\affiliation{Inter-University Institute for Data Intensive Astronomy, and Department of Physics and Astronomy, University of the Western Cape, Robert Sobukwe Road, 7535 Bellville, Cape Town, South Africa\\}

\author[0000-0002-4117-343X]{Jiajun Zhang}
\affiliation{Shanghai Astronomical Observatory, Chinese Academy of Sciences, Shanghai 200030, China\\}

\author[0000-0003-0631-568X]{Yougang Wang}
\affiliation{National Astronomical Observatories, Chinese Academy of Sciences, Beijing 100101, China\\}
\affiliation{School of Astronomy and Space Science, University of Chinese Academy of Sciences, Beijing 100049, China\\}

\correspondingauthor{Xin Zhang}
\author[0000-0002-6029-1933]{Xin Zhang}
\email{zhangxin@mail.neu.edu.cn}
\affiliation{Key Laboratory of Cosmology and Astrophysics (Liaoning)\\
\& College of Sciences, Northeastern University, Shenyang 110819, China\\}
\affiliation{National Frontiers Science Center for Industrial Intelligence and Systems Optimization, Northeastern University, Shenyang 110819, China}
\affiliation{Key Laboratory of Data Analytics and Optimization for Smart Industry (Ministry of Education), Northeastern University, Shenyang 110819, China}

\correspondingauthor{Xuelei Chen}
\author[0000-0001-6475-8863]{Xuelei Chen}
\email{xuelei@cosmology.bao.ac.cn}
\affiliation{Key Laboratory of Cosmology and Astrophysics (Liaoning)\\
\& College of Sciences, Northeastern University, Shenyang 110819, China\\}
\affiliation{National Astronomical Observatories, Chinese Academy of Sciences, Beijing 100101, China\\}
\affiliation{School of Astronomy and Space Science, University of Chinese Academy of Sciences, Beijing 100049, China\\}



\begin{abstract}
The peaked-spectrum (PS) sources exhibit turnover characteristics in their broad radio spectra. However, the mechanism underlying this phenomenon remains elusive.
The two most common hypotheses are synchrotron self-absorption (SSA) and free-free absorption (FFA). 
By incorporating multiple absorption scenarios, we propose a multi-mechanism hybrid (MMH) model, which aligns well with current observational data and provides a good physical explanation. Using the GLEAM survey data, we identified a sample of $4,315$ sources with peak frequencies approximately between $72$--$3000$ MHz, most of which are MHz-peaked-spectrum sources (MPS). 
Our analysis shows that instead of SSA, the FFA is the dominant mechanism in producing the spectral turnover for most of the sources in this sample. The index of the optically thick spectrum $\alpha_{\rm thick}$ has a lower boundary due to the FFA, and the steeper $\alpha_{\rm thick}$ indicates a complex multi-absorption mechanism. In particular, the external FFA produces substantial $\alpha_{\rm thick}$, which exhibits a weak correlation with the peak frequency. Future ultra-long wavelength observations would also provide data on the spectrum of these sources at even lower frequencies. Determining the absorption mechanism that shaped the spectrum of these sources would be a crucial part of understanding their nature.

\end{abstract}




\section{Introduction}
\label{sec:intro}

Peaked-spectrum (PS) sources are a distinctive subset of active galactic nuclei (AGN), characterized by significant turnovers in their radio spectra. These turnovers typically occur within the frequency range of several \GHz to a few hundred \MHz. Based on their peak frequencies in the observer’s frame, these sources are further classified into High-Frequency-Peaked (HFP) sources \citep{Dallacasa2000}, GHz-Peaked-Spectrum (GPS) sources \citep{GopalK1983, Stanghellini1998}, and MHz-Peaked-Spectrum (MPS) sources \citep{Coppejans2015, Coppejans2016a, Coppejans2016b}, which have peaks above a few \GHz, around \GHz, and around a few hundred of \MHz, respectively.

In terms of morphology, PS sources are often classified as Compact-Steep-spectrum radio Sources (CSS) \citep{Fanti1990}, also called Medium-Symmetric Objects \citep[MSO, ][]{fanti1995}, and Compact-Symmetric Objects \citep[CSO, ][]{Wilkinson1994, Peck2000, Kiehlmann2024}. These sources exhibit a wide range of physical scales, from a few \pc to about $20$ \kpc. They are classified primarily on the basis of their physical scales, with CSSs generally having linear scales greater than 1 \kpc, while CSOs are smaller. 
Collectively, these differently classified sources are referred to as PS sources \citep{ODea1998, ODea2021}. 
Generally, CSS and MPS are related, while CSO and GPS/HFP are related, merely different classifications for the same batch of sources. When considering the effect of redshift, the inverse relation between their linear size and the peak frequency of the rest frame is referred to as the linear size turnover relation (hereafter LTR) for PS sources \citep{ODea1997, ODea1998, Jeyakumar2016}.

Studying PS sources is crucial for understanding the early stages of radio galaxy evolution and the physical processes of their radio emission. The turnover of the radio spectrum at low frequencies is believed to be due to absorption, and two primary absorption mechanisms are believed to be likely responsible: synchrotron self-absorption (SSA) and free-free absorption (FFA) \citep{ODea1998, Tingay2003, Marr2014, Callingham2015, ODea2021}. The action of these mechanisms depends on the physical conditions inside and around the radio sources and may be used to derive deeper insights and interesting applications on these sources. For example, if the empirical relationships of LTR are confirmed, observations of more PS sources with future ultra-long wavelength instruments may provide new cosmological standard rulers.

Two main hypotheses have been proposed to explain these spectral turnovers, which are based on the two absorption mechanisms. The ``youth" hypothesis posits that different types of PS sources represent various evolutionary stages of radio galaxies, from the youngest HFP to GPS/CSO, then evolving into MPS/CSS, and eventually becoming Fanaroff-Riley type I (FR I) or II (FR II) galaxies \citep[e.g.][]{Fanaroff1974, KunertB2010, An2012}. According to this hypothesis, the primary absorption mechanism that causes spectral turnover is SSA. However, this hypothesis results in an overabundance of PS sources relative to large AGNs and cannot alone explain all PS sources \citep{ODea1997, Snellen2000, An2012}.
Recently, \cite{Kiehlman2024} suggests that only about one-fifth of edge dimmed CSO objects are likely to develop into large-scale radio sources observed today. In contrast, most edge brightened CSOs are unlikely to evolve into larger structures and are instead classified as short-lived rather than ``young".
Although this hypothesis requires further studies, it provides an explanation for a large number of small sources without resorting to a dense medium that frustrates their growth and is responsible for FFA.

The ``frustration" hypothesis suggests that these sources are in dense environments where interactions with the surrounding dense gas lead to FFA being the dominant mechanism for spectral turnovers \citep[e.g.][]{van1984, Bicknell1997, Dicken2012, Callingham2015, Bicknell2018}. Traditionally, the relationships between the radio power and linear size of the PS sources and the LTR were believed to be hallmarks of the ``youth" hypothesis. However, the frustration hypothesis can also produce similar results \citep[e.g.][]{Bicknell1997, Keim2019, Curran2024}.

Distinguishing between these absorption mechanisms is essential for understanding the internal and surrounding physical environments of radio sources and the evolutionary processes of AGNs. The preferred approach involves finely sampling the optically thick region of the radio spectrum below the peak frequency. By analyzing the spectral index in the optically thick region, different absorption mechanisms can be distinguished to some extent \citep[e.g.][]{Snellen2009, Callingham2017}.

Low frequency surveys such as the GaLactic and Extragalactic All-sky Murchison Widefield Array \citep[GLEAM;][]{Wayth2015} of the Murchison Widefield Array \citep[MWA;][]{Tingay2013}, and the LOFAR Two-Metre Sky Survey \citep[LoTSS;][]{Shimwell2022} and the LOFAR LBA Sky Survey \citep[LoLSS;][]{deG2021} by the LOw-Frequency ARray \citep[LOFAR;][]{vanH2013} provide ample low frequency radio source data, covering much of the optically thick region for many GPS/HFP sources \citep[e.g.][]{Callingham2017, Slob2022, Ballieux2024}. 

The different models of spectral turnover can be distinguished by performing a comprehensive spectral fitting on the PS sources. 
The SSA mechanism is expected to work only within the source. On the other hand, FFA can be divided into different types depending on its location and uniformity. Other mechanisms may also be at work, e.g., spectral aging \citep[e.g.][]{Turner2018, Quici2021}, inverse-Compton losses \citep[e.g.][]{Potter2013, klein2018}, the Razin-Tsytovich effect in plasma \citep[e.g.][]{Melrose1980, Dougherty2003, ravi2019}, bremsstrahlung \citep[e.g.][]{klein2018}, ionization losses \citep[e.g.][]{murphy2009, Basu2015, mckean2016},  resulting in various model combinations. Among these, spectral aging and inverse-Compton losses cause the radio spectrum to steepen at high frequencies, while SSA, FFA, the Razin-Tsytovich effect, and ionization losses flatten the spectrum at low frequencies. The peaked spectrum may be the result of a combination of these effects.

In \cite{Callingham2015}, the goodness of fit for different models is used to compare models with different absorption mechanisms. A broadband measurement of the known PS source PKS B0008-421 is made, and nine different models are used to fit the observed spectral data, to understand the absorption mechanisms causing the spectral turnovers. Unfortunately, none of these models fit well, even the relatively well-performing ones—the model with inhomogeneous FFA \citep{Bicknell1997} and spectral aging, and the two-component SSA model, required extreme physical conditions, such as very high electron column densities ($\gtrsim 10^{20}\,\rm cm^{-2}$) or magnetic field strengths ($\sim 4.1\,\rm Gauss$). Subsequent studies mainly used general curve models that lack explicit physical explanations \citep{Callingham2017, Keim2019, Slob2022, He2024}. However, these models are currently used individually, without considering the combined effects of SSA and FFA.

In this work, we propose a new multimechanism hybrid (hereafter MMH) spectral model for PS sources and inspect the fitting goodness with current observational data. This paper is structured as follows. Section~\ref{sec:A Hybrid Spectral Model} describes the construction of the MMH model. Section~\ref{sec:data sample} provides an overview of the observational data that we used. In Section~\ref{sec:Selectio of PS sources}, we discuss the selection criteria used to identify PS sources. Section~\ref{sec:results} presents the results of applying the model to the PS source sample. Section~\ref{sec: Discussion} discusses several different absorption mechanisms and the impact of future ultra-long wavelength observations on studying PS sources. We adopted the Planck2018 standard Lambda cold dark matter ($\Lambda$CDM) cosmological model, with the following parameters: $\Omega_{\rm M} = 0.31$, $\Omega_\Lambda = 0.69$, and the Hubble constant $H_0=67.66\  {\rm km}\,{\rm s}^{-1}\,\Mpc^{-1}$ \citep{Planck2020}.

\section{A Multi-Mechanism Hybrid Spectral Model}
\label{sec:A Hybrid Spectral Model}
\begin{figure}
\centering
\includegraphics[width=\columnwidth]{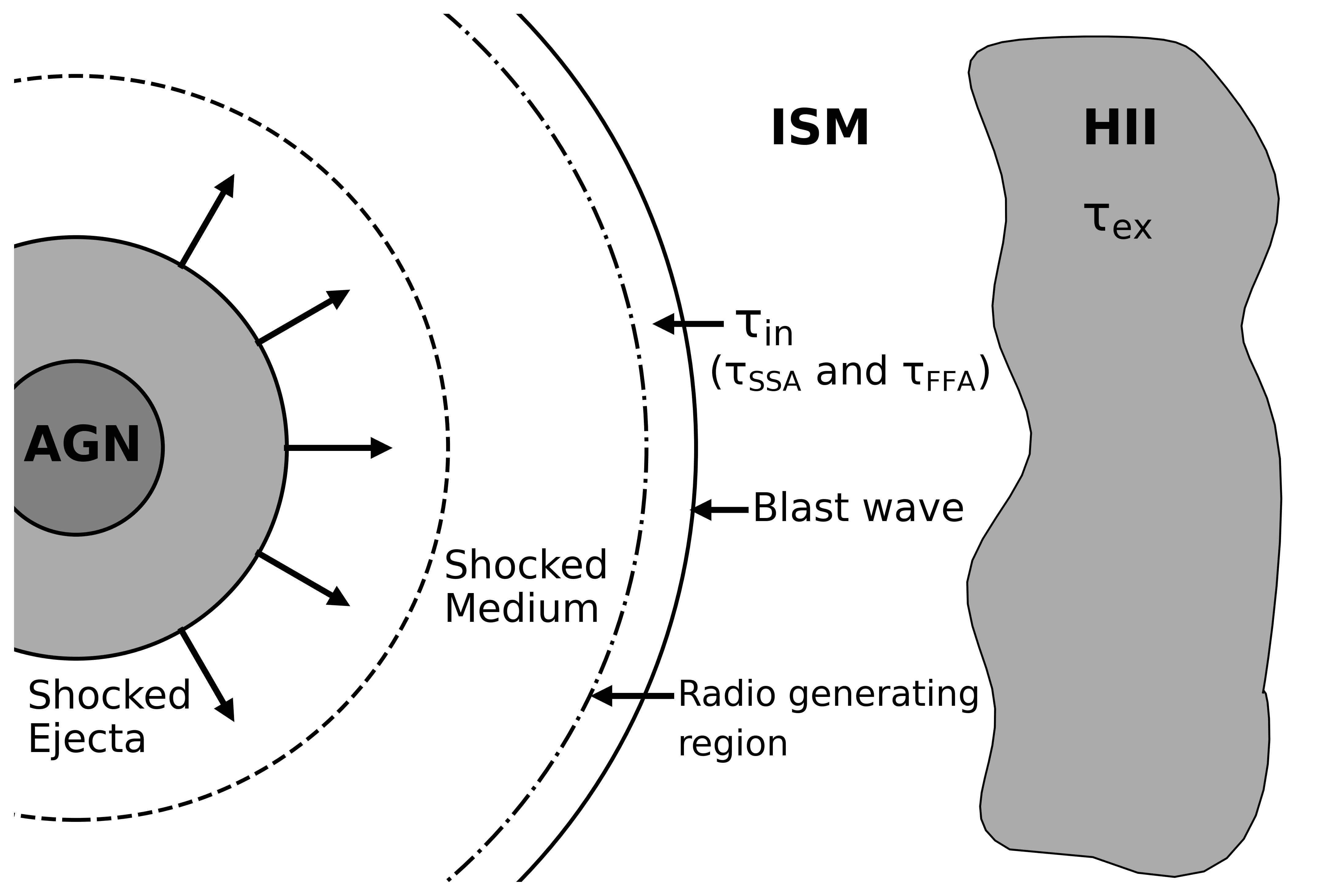}
\caption{
Schematic representation of AGN radio emission and absorption mechanisms. The diagram illustrates key components, including the central AGN, interstellar medium (ISM), and distant ionized hydrogen (HII) gas, as well as their roles in shaping the radio spectrum. Synchrotron emission originates near the explosion wavefront (core or jet lobes), while multiple absorption components (e.g., internal SSA, internal FFA, and external FFA) act at different scales to modify the observed spectrum. This visual highlights the physical locations and contributions of the mechanisms incorporated into the MMH model. Note: Proportions are not representative.
}
\label{fig:cartoon}
\end{figure}

It is widely accepted that bright extragalactic PS sources are part of the evolutionary stage of radio AGNs, the radiation mainly arises from non-thermal synchrotron radiation, with debates focusing on specific absorption mechanisms \citep{Bicknell2018, ODea2021, Nascimento2022, Ross2023}. 
Our model of the PS source is inspired by the supernova radiation model \citep{Weiler2002}, which essentially combines synchrotron radiation emission with multiple absorption mechanisms, 
including both the SSA and the internal FFA within the emission region and the external FFA outside the emission region, respectively.
Unlike the supernova model, we neglect the local circumstellar medium, which has a negligible effect on the sources. We also neglect the time evolution parameters, as our observations usually do not show a significant evolution of the source within the observation time scale \citep{Bell2014, Ross2021, Ross2022}. These simplifications allow us to fit the data, which typically includes 20 available frequency data points per source. Figure~\ref{fig:cartoon} shows a cartoon diagram of our model, illustrating the positions of different absorptions used in the subsequent discussion. The actual internal structure of an AGN is usually very complex \citep[e.g.][]{Keim2019, Brienza2021}; regions near the jet lobes and core may be affected by shocks, producing non-thermal radio emission. However, since most PS sources are at present unresolved, a spherical approximation is sufficient. 

Consequently, a straightforward MMH model of the PS source spectrum, incorporating multiple absorption mechanisms, can be expressed as \citep{Weiler2002}
\begin{equation}
    \left(\frac{S_\nu}{{\rm Jy}}\right) = \left( \frac{K}{\Jy} \right) \left( \frac{\nu}{\MHz} \right) ^{\alpha_{\rm thin}} \left( \frac{1-e^{-\tau_{\rm in}}}{\tau_{\rm in}} \right) e^{-\tau_{\rm ex}},
	\label{eq:PSmodel}
\end{equation}
where $S_\nu$ represents the flux density at frequency $\nu$, $K$ denotes the normalization constant of flux density, $\alpha_{\rm thin}$ is the spectral index of the synchrotron emission, which corresponds to the spectral index in the optically thin spectrum ($\tau \ll 1$). The parameter $\tau_{\rm ex}$ describes the attenuation caused by the absorption of the medium located away from the source. Assuming the opacity depends on thermal, ionized hydrogen (HII) with a constant electron density ($n_{\rm e}$) and electron temperature ($T_{\rm e}$), a good approximation is \citep{Mezger1967, Condon2016, Cong2021}:
\begin{equation}
    \tau_{\rm ex} \approx 0.654 \left( \frac{T_{\rm e}}{\rm 10^4 \, K} \right) ^{-1.35} \hspace{-0.15cm} \left( \frac{\nu}{\MHz} \right) ^{-2.1} \left( \frac{\rm EM}{\pc \, \rm cm^{-6}} \right),
	\label{eq:ioniz}
\end{equation}
where the emission measure EM represents the integral of $n_{\rm e}^2$ along the line of sight path distance $s$:
\begin{equation}
    \left(\frac{\rm EM}{\pc \,\rm cm^{-6}}\right) = \int \left( \frac{n_{\rm e}}{\rm cm^{-3}} \right) ^{2} \left( \frac{{\rm d}s}{\pc} \right).
	\label{eq:em}
\end{equation}
For simplicity, we define:
\begin{equation}
\begin{split}
    \tau_{\rm ex} = \left( \frac{\nu}{\nu_{\rm ex}} \right) ^{-2.1},
	\label{eq:taudis}
\end{split}
\end{equation}
where $\nu_{\rm ex}$ denotes the observed frequency at which the optical depth influenced by the external homogeneous medium $ \tau_{\rm ex} \sim 1$, assuming the peak frequency at the source's rest frame is $\nu_{\rm source}$, so $\nu_{\rm ex} = \nu_{\rm source}\rm (1+z)^{-1}$. Additionally, assuming the $n_{\rm e}$ of the HII region along the line of sight is constant within the scale $s$, $\nu_{\rm ex}$ can also be expressed as:
\begin{equation}
\begin{split}
    \left(\frac{\nu_{\rm ex}}{\rm MHz}\right) \approx 0.817 \left( \frac{T_{\rm e}}{\rm 10^4\,K} \right)^{-0.64}\hspace{-0.15cm} \left( \frac{n_{\rm e}}{\rm cm^{-3}} \right) ^{0.95} \left( \frac{s}{\pc} \right)^{0.48}\hspace{-0.2cm} (1+z)^{-1}.
	\label{eq:nudis}
\end{split}
\end{equation}
Hereafter, we use ``external" or ``external FFA" to refer to this absorption mechanism.

In addition, the internal absorption mechanism may not be adequately described by either SSA or FFA alone. We assume that the observed spectrum results from the combined effects of both mechanisms. Therefore, the internal absorption $\tau_{\rm in}$ term in Equation~(\ref{eq:PSmodel}) can consist of these two components:
\begin{equation}
    \tau_{\rm in} = \tau_{\rm SSA} + \tau_{\rm FFA},
	\label{eq:tau_int}
\end{equation}
where the optical depth produced by SSA ($\tau_{\rm SSA}$) is:
\begin{equation}
    \tau_{\rm SSA} = \left( \frac{\nu}{\nu_{\rm SSA}} \right) ^{\alpha_{\rm thin} -2.5},
	\label{eq:tau_ssa}
\end{equation}
and thermal FFA ($\tau_{\rm FFA}$) is:
\begin{equation}
    \tau_{\rm FFA} = \left( \frac{\nu}{\nu_{\rm FFA}} \right) ^{-2.1}.
	\label{eq:tau_int}
\end{equation}
Here, $\nu_{\rm SSA}$ and $\nu_{\rm FFA}$ represent the characteristic frequencies when considering only the SSA or FFA cases, respectively. For $\nu_{\rm SSA}$, we have \citep{Kellermann1981, Tingay2003, Keim2019}:
\begin{align}
    \left(\frac{\nu_{\rm SSA}}{\rm MHz} \right) & \approx f(\alpha_{\rm thin}) \times 10^3 \nonumber\\
    &\times \left( \frac{ B}{\rm Gauss} \right) ^{1/5} \vspace{-0.15cm} \left( \frac{S_{\rm SSA}}{\rm Jy
    } \right) ^{2/5} \left( \frac{\theta}{\rm mas} \right) ^{-4/5}\vspace{-0.3cm}  (1+z)^{1/5},
	\label{eq:nu_ssa}
\end{align}
where $B$ represents the magnetic field strength, and $\theta$ denotes the angular size of the source. $f(\alpha_{\rm thin})=10/b^{1/5}(\alpha_{\rm thin})$ is estimated based on the Table 1 calculated by \cite{Marscher1983} (See also analytical derivation in \cite{Pushkarev2019}), with a range of approximately 7 to 10, and it is not very sensitive to $\alpha_{\rm thin}$. Therefore, in this paper, we use an approximate value of 8.1 \citep[e.g.][]{Callingham2015, Ross2023}. Additionally, $S_{\rm SSA}$ refers to the peak flux density caused by SSA absorption. However, since our model also considers the impact of FFA absorption, the peak frequency $\nu_{\rm p}$ is typically not equal to $\nu_{\rm SSA}$. We approximate $S_{\rm SSA}$ using a simple linear relationship:
\begin{equation}
    S_{\rm SSA} = S_{\rm p} \left( \frac{\nu_{\rm SSA}}{\nu_{\rm p}} \right) ^{\alpha_{\rm thin}}.
	\label{eq:S_ssa}
\end{equation}
Note that the $S_{\rm p}$ and the $\nu_{\rm p}$ are determined by fitting the actual shape of the model curve. Thus, the only none-observable quantity in Equation~(\ref{eq:nu_ssa}) is the magnetic field strength $B$, which can be derived as:
\begin{align}
    \left(\frac{B}{\rm Gauss}\right) &\approx 2.868 \times 10^{-20} \left( \frac{\nu_{\rm SSA}}{\rm MHz} \right)^{5-2\alpha_{\rm thin}} \nonumber\\
    &\times \left( \frac{\nu_{\rm p}}{\rm MHz} \right)^{2\alpha_{\rm thin}} \left( \frac{S_{\rm p}}{\rm Jy} \right)^{-2} \left( \frac{\theta}{\rm mas} \right)^4 (1+z)^{-1}.
	\label{eq:B}
\end{align}
This formula is used to estimate the internal magnetic field strength $B$ of sources based on the fitting results. For $\nu_{\rm FFA}$, similar to $\nu_{\rm ex}$, we have: 
\begin{equation}
    \left(\frac{\nu_{\rm FFA}}{{\rm MHz}}\right) \approx 0.817 \left( \frac{T_{\rm e^{\prime}}}{10^4 \, \rm K} \right)^{-0.64}  \hspace{-0.15cm}\left( \frac{n_{\rm e^{\prime}}}{\rm cm^{-3}} \right)^{0.95} \left( \frac{s^{\prime}}{\rm pc} \right) ^{0.48}\hspace{-0.15cm} (1+z)^{-1}.
	\label{eq:nuffa}
\end{equation}
The notation ($^{\prime}$) is used to distinguish these variables from those in 
Equation~(\ref{eq:nudis}), though $s^{\prime}$ here represents the scale of the emission region. Hereafter, we use ``FFA" or ``internal FFA" to refer to this absorption mechanism. 
We recognize that the hot thermal plasma responsible for internal FFA and the relativistic electrons responsible for SSA may have different physical scales. However, both arise from the physical scale associated with the AGN and therefore should not differ too much. High-resolution polarization measurements would be required to effectively distinguish between them. Given that our model is a simplified one, we assume that their emission regions share the same spatial scale:

\begin{equation}
    \left(\frac{s^{\prime}}{\pc}\right) 
    = 4.848 \times 10^{-9} \left( \frac{D_{\rm A}}{\rm pc} \right) \left( \frac{\theta}{\rm mas} \right),
	\label{eq:sourcesize}
\end{equation}
where $D_{\rm A}$ is angular diameter distance. 

The MMH model has five free parameters: $K$, $\nu_{\rm ex}$, $\alpha_{\rm thin}$, $\nu_{\rm SSA}$, and $\nu_{\rm FFA}$. This model introduces only one more free parameter compared to the general curve model \citep[e.g.][]{Callingham2017, He2024}. For clarity, the Equation~(\ref{eq:PSmodel}) described by these five free parameters is as follows:
\begin{align}
    \left(\frac{S_\nu}{\rm Jy}\right) & = \left( \frac{K}{\Jy} \right) \left( \frac{\nu}{\MHz} \right) ^{\alpha_{\rm thin}}  \nonumber\\
    & \times \left( \frac{1-e^{-(\nu / \nu_{\rm SSA})^{\alpha_{\rm thin}-2.5} -(\nu / \nu_{\rm FFA})^{-2.1}}}{(\nu / \nu_{\rm SSA})^{\alpha_{\rm thin}-2.5} +(\nu / \nu_{\rm FFA})^{-2.1}} \right) e^{-(\nu / \nu_{\rm ex})^{-2.1}}.
	\label{eq:pspmodel}
\end{align}

\section{Data}
\label{sec:data sample}
Identifying PS sources and fitting their spectra requires dense frequency point data and sufficiently wide spectral line bandwidths. The primary data used in this study comes from the GLEAM survey. Moreover, we include observational data from different frequency bands, for example, the VLITE Commensal Sky Survey \citep[VCSS, \,][]{Peters2021}, the Sydney University Molonglo Sky Survey \citep[SUMSS,\,][]{Bock1999, Mauch2003}, the Rapid ASKAP Continuum Survey \citep[RACS,\,][]{McConnell2020, Hale2021}, the NRAO VLA Sky Survey \citep[NVSS,\,][]{Condon1998}, and the Very Large Array Sky Survey \citep[VLASS,\,][]{Lacy2020}. We summarize these surveys, along with the important parameters relevant to this study, in Table~\ref{tab:catalog}.

We employed four high-frequency surveys, SUMSS, RACS, NVSS, and VLASS, which collectively cover the entire sky region observed by GLEAM, offering nearly an order of magnitude improvement in sensitivity. 
This study is the first to combine data from VLASS and its companion project VCSS with GLEAM for the study of PS sources. The excellent sensitivity and resolution of VLASS allow us to avoid relying on unresolved sources in GLEAM observations.

\subsection{GLEAM}
\label{sec:gleam}
The GLEAM survey is a comprehensive low-frequency radio survey conducted using the MWA located in Western Australia. The GLEAM survey covers the frequency range from 72 to 231 \MHz, capturing detailed images of both Galactic and extragalactic radio sources \citep{Wayth2015}. The MWA consists of 128 antenna tiles. Each tile contains 16 dual-polarization dipoles, offering a large field of view and high sensitivity. This design allows the MWA to cover over 30,000 square degrees of the sky, including the entire southern hemisphere, with high angular resolution and sensitivity \citep{Tingay2013}. It is currently the most comprehensive low-frequency radio observation available. The survey has produced a significant amount of data, resulting in the creation of high-quality radio images. These datasets have facilitated numerous studies, enhancing our understanding of the radio sky. 

\subsection{VLASS and VCSS}
\label{sec:vlass and vcss}
VLASS covers the sky north of $-40^\circ$ declination in the S-band (2–4 \GHz), making it one of the highest-resolution all-sky radio surveys available. The survey aims to map the entire observation area across three distinct epochs, with the first epoch completed in 2019. VLASS offers unparalleled resolution and sensitivity compared to the other surveys listed in Table~\ref{tab:catalog}, making it a powerful tool for studying the radio sky. 

One key application of VLASS data is the search for deep radio AGNs (DRAGN catalog), as emphasized in \cite{Gordon2023}. The DRAGN catalog focuses on using the high-resolution capabilities of VLASS to identify AGNs with substantial radio emissions. By cross-matching with other surveys and incorporating spectral analysis, the DRAGN catalog has enhanced our understanding of the AGN population.

VLASS has completed the data release of two epochs under its ``Quick Look" mode. However, we noted some issues in epoch 1, such as the systematic underestimation of flux density, astrometry problems, and ghost image in bright sources. These issues have been addressed in epoch 2 observations. Therefore, we used the DRAGN catalog based solely on the VLASS epoch 2 calculations.

The VCSS is a sub-1 \GHz survey utilizing data collected at 340 \MHz from the VLA during the same observations conducted for VLASS. The VCSS uses the VLITE system (VLA Low-band Ionosphere and Transient Experiment), a commercial system operating at low frequencies, which has been collecting data since 2014. This survey provides a significant contribution to ongoing radio surveys by extending the observed frequency range of the VLA, offering higher resolution and sensitivity in the low-frequency regime.

VCSS observations were conducted simultaneously with VLASS epoch 1 observations. As a result, the aforementioned issues in VLASS epoch 1 also impacted VCSS. However, the catalog provides flux density corrections, and we used the corrected flux densities in our analysis.

\subsection{Additional surveys}
\label{sec:Additionalsurveys}

In addition to the aforementioned surveys, we also included several other all-sky surveys, such as the VLA Low-Frequency Sky Survey Redux \citep[VLSSr,\,][]{Lane2014}, the Molonglo Reference Catalogue \citep[MRC,\,][]{Large1981, Large1991}, and the TIFR Giant Metrewave Radio Telescope (GMRT) Sky Survey \citep[TGSS,\,][]{Interma2017}. The relevant parameters of these surveys are shown in the lower rows of Table~\ref{tab:catalog}.

Compared with other surveys, we found that these datasets exhibit systematic errors, with observed flux densities often deviating significantly from expected values. 
Following the previous analysis by \citet{Callingham2017}, such additional surveys 
were only used for cross-checking with our selected PS sources but not for the spectral model fitting 
in this work.

\subsection{Systematic Uncertainties}
\label{sec:systematic uncertainties}

In this study, we fit the MMH model with the spectra data from GLEAM, VCSS, SUMSS, RACS, NVSS, and VLASS to identify PS sources. These surveys are carried out at different times, utilize various calibration sources, and exhibit different resolutions and sensitivities. Consequently, there are inevitably systematic errors of varying degrees between datasets. Based on recommendations from the literature detailed below, we applied approximately three times the suggested systematic errors to reflect our concerns about the reliability of the data.

We incorporated internal systematic uncertainties into the GLEAM catalog, assigning $6\%$ to sources within the range $-72^\circ < \delta < 18.5^\circ$ and $9\%$ to sources outside this range \citep{Hurley-Walker2017}. Furthermore, we applied a systematic error of $20\%$ to VCSS \citep{Peters2021} and of $10\%$ to each of SUMSS \citep{Bock1999, Mauch2003}, RACS \citep{Hale2021}, NVSS \citep{Condon1998}, and VLASS \citep{Lacy2020, Gordon2021}. The final uncertainty for each source combines the catalog error ($\sigma_{\rm rms}$) and the added systematic error ($\sigma_{\rm sys}$), represented as $\sigma = \sqrt{\sigma_{\rm rms}^2 + \sigma_{\rm sys}^2}$.

\begin{table*}
	\centering
	\caption{A summary of recent radio discrete source catalog covering extensive sky areas and different frequency bands. 
    The measurements of the first six surveys are adopted in this work for spectral modeling. 
    The last three catalogs are assisted in verifying the accuracy of the existing spectral fits. 
    The publication year refers to when the first batch of survey data was published instead of the final catalog release date. For some observations, angular resolution varies with declination, we list only the highest resolution achieved under general observation conditions. The flux density limit is defined similarly to sensitivity, typically $5$ times the rms. $N_{\rm Isolated}$ represents the number of sources identified as \textit{isolated} when cross-matched with GLEAM in these surveys. $\sigma_{\rm sys}$ represents the systematic error, with only the surveys used in the fitting process annotated. The related references are cited in Section~\ref{sec:data sample}.}
	\label{tab:catalog}
	\begin{tabular}{lcccccccr}
		\hline
		Survey & Data published & Frequency & Survey region & Resolution & Flux density limit & Bandwidth & $N_{\rm Isolated}$ & $\sigma_{\rm sys}$\\
        &        &     (\MHz) &   & (arcsec)  &   $\rm (mJy\ beam^{-1})$ & (\MHz)& &\\
		\hline

        GLEAM & 2016 & 72-231& $\delta < +30^\circ$   & 120 & 30 & $4 \times 7.68^{*}$ & 225909 & $2-3\%$\\
        VCSS  & 2021 & 340 & $\delta > -40^\circ$ & 15 & 200$^{**}$ & 33.6 & 15616 & $\sim 10\%$\\
        SUMSS & 2003 & 843   & $\delta < -30^\circ$  & 45 & 6 & 3 & 83224 & $3\%$\\
        RACS  & 2020 & 887.5 & $-85^\circ<\delta<+30^\circ$   & 25 & 3 & 288 & 224255 & $2.5\%$\\
        NVSS  & 1998 & 1400  & $\delta > -40^\circ$  & 45 & 3 & 50 & 166191  & $2\%$\\
        VLASS & 2020 & 3000 & $\delta > -40^\circ$   & 2.5 & 0.35 & 2000 & 86902 & $3\%$\\
        \hline
        VLSSr & 2007 & 74    & $\delta > -30^\circ$   & 75 & 350 & 1.56 & 24025 & \\         TGSS  & 2016 & 150 & $\delta > -53^\circ$ & 25& 20 & 16.7 & 153067 & \\
        MRC$^{***}$   & 1981 & 408   & $-85^\circ < \delta < +18.5^\circ$ & 150 & 700 & 2.5 & 5353  & \\
        \hline
        \multicolumn{9}{l}{$^{*}$ The 30.72 \MHz bandwidth is subdivided into four 7.68 \MHz sub-channels, so the actual bandwidth of the data at each frequency} \\
        \multicolumn{9}{l}{ point is 7.68 \MHz.}\\
        \multicolumn{9}{l}{$^{**}$ This corresponds to approximately $50$ times of noise rms instead of 5 times.} \\
        \multicolumn{9}{l}{$^{***}$ The MRC survey region is expressed in J1950 coordinates.}\\
        \hline
	\end{tabular}
\end{table*}

\subsection{Cross-matching Routine}
\label{sec:Cross-matching Routine}

We used The Positional Update and Matching Algorithm \citep[PUMA,\,][]{puma2017} to perform cross-matching between GLEAM and VCSS, SUMSS, RACS, NVSS, and VLASS. PUMA provides a cross-matching approach designed to align and match astronomical data from different catalogs. It integrates both positional and spectral information to ensure robust matching, making it particularly useful in cases where simple positional cross-matching might lead to ambiguities or errors, such as when sources are closely spaced or when catalog resolutions differ significantly. The core of PUMA’s matching process is based on a Bayesian framework, which calculates the posterior probability that two sources from different catalogs represent the same astrophysical object. This approach allows the algorithm to scale effectively when handling multiple catalogs, ensuring reliable results even with large data sets.

In this work, the GLEAM catalog was used as the base catalog for cross-matching with VCSS, SUMSS, RACS, NVSS, and VLASS. The matching angular radius was set to the FWHM of the MWA broadband images, approximately 2\arcmin20\arcsec. We set the positional probability threshold for cross-matching at 0.99 and only accepted sources classified as \textit{isolated} by PUMA.  This means that there is only one possible cross-match, it is accepted directly if the positional probability is above a threshold. As a result, 225,909 sources were selected from the 307,455 sources in the GLEAM catalog. The number of these sources cross-matched with other surveys is listed in Table~\ref{tab:catalog}.

\section{Identification of PS sources}
\label{sec:Selectio of PS sources}
Not all of these 225,909 isolated sources selected from the GLEAM catalog have the shape of the peaked spectrum within the observed frequency ranges.
Most radio sources approximate a power-law \citep{Callingham2017, He2024}. 
Additionally, some sources exhibit peculiar and erratic spectral shapes, which could be due to observational errors, weather conditions, source variability, or the superposition of spectra from multiple sources within the resolution limitations. 
These sources could introduce biases in subsequent analyses and are excluded to prevent
contamination of the PS sample. 
The data quality selection criteria are set as follows.

(i) A source must have more than $8$ frequency measurements from the GLEAM catalog
with signal-to-noise ratio (${\rm SNR} >3$). 

(ii) A minimal flux density elimination is applied to provide a reliable sample
of peaked-spectrum sources. 
Based on the wideband images, the completeness of the GLEAM extragalactic catalog is
$\sim 90\%$ with a flux density limit of 0.16 \Jy \citep{Hurley-Walker2017}. 
In this work, we only select sources with $S_{\rm 200\MHz, wide}>0.16\, \Jy$, 
where $S_{\rm 200\MHz, wide}$ represents the flux density in the wideband image. 

(iii) A source must have at least one frequency measurement available in the high-frequency surveys, i.e. SUMSS, RACS, NVSS, or VLASS.

After applying these criteria, there are $97,976$ sources remaining. We then perform the fit of the model on this sample.
We utilize the Python package \texttt{emcee} \citep{emcee2013} and the Markov Chain Monte Carlo (MCMC) algorithm to sample the posterior probability density functions of each model parameter, 
assuming a Gaussian likelihood function. 
The least squares method is used to estimate an initial value, followed by sampling under physically reasonable uniform priors. 

The PS sources are selected according to the goodness of fit. We fit the observation data both with our MMH model and a power-law (PL) model,
\begin{equation}
    \left(\frac{S_\nu}{{\rm Jy}}\right) = \left( \frac{A}{\Jy} \right) \left( \frac{\nu}{\rm MHz} \right) ^\alpha,
	\label{eq:powerlaw}
\end{equation}
where $A$ characterizes the amplitude and $\alpha$ is an arbitrary spectral index. 
The peaked-spectrum shape is indicated if the goodness of fit is significantly improved with the MMH model. 

We employ two statistical methods, i.e. the F-test and the Bayesian Information Criterion (BIC), 
to compare the fitting goodness of different spectral models. 
These methods provide reliable means to evaluate whether there is a significant improvement in fit between different models and to balance model fitness with model complexity.

The F-test is used to compare two nested models, where one model is a special case of the other. This test determines whether the additional parameters in the more complex model significantly improve the fit of the model to the data. The test statistic, $F$, is calculated as follows:
\begin{equation}
    F = \frac{(\chi^2_1 - \chi^2_2) / (k_2 - k_1)}{\chi^2_2 / (n - k_2)},
	\label{eq:F}
\end{equation}
\begin{equation}
    \chi^2_{i\,(i=1,\ 2)}=\sum_{n} \frac{(p_n^{\rm data}-p_{n,\ i}^{\rm mod})^2}{\sigma_n^2},
	\label{chi2}
\end{equation}
where $n$ is the number of measurements, $p_n^{\rm data}$ denotes the observed spectrum, 
and $\sigma_n$ is the measurement uncertainties. $p_{n,\ i}^{\rm mod}$ represent the $i$-th model, that is,
$i=1$ denotes the simple power-law (PL) model and $i=2$ for the MMH model, respectively.
$k_1$ and $k_2$ are the numbers of parameters defined in the power-law and MMH models. 
The F-statistic follows an F-distribution with $(k_2 - k_1, n - k_2)$ degrees of freedom. 
A p-value is derived from the F-statistic, which indicates whether the improvement in fit is statistically significant \citep{Akaike1974}.

The BIC is another model selection criterion that incorporates both the goodness of fit and the complexity of the model. BIC is calculated using the following formula:
\begin{align}
    \text{BIC} = -2\, \ln \, \mathcal{L}_{\rm max} + k\, \ln\, n,
	\label{eq:bic}
\end{align}
where $\mathcal{L}_{\rm max}$ is the maximum likelihood of the model. 
The BIC penalizes models with more parameters, thus discouraging overfitting. 
The model with the lowest BIC is considered the best among the candidates. 
This criterion balances the trade-off between model complexity and goodness of fit, providing a more parsimonious model selection framework \citep{Schwarz1978}.

The criteria used for selecting PS sources are detailed below:

(i) The sources that pass the quality selection are then applied to the F-test.
The sources with a p-value less than 0.04 are accepted as PS source candidates.
If the p-value is between 0.04 and 0.06, we further inspect the differences in BIC
$\Delta {\rm BIC} = \text{BIC}_{\rm PL} - \text{BIC}_{\rm MMH}$
where $\text{BIC}_{\rm MMH}$ and $\text{BIC}_{\rm PL}$ are the BIC values calculated from the MMH model and the PL model. If $\Delta {\rm BIC}  > 1.5$, the sources are also taken as PS source candidates. All sources with p-values greater than $0.06$ are considered to favor the PL model and are excluded.

(ii) We then exclude sources with $\chi^2_{\rm MMH} > 2$ to avoid those with erratic shapes. This simple approach effectively eliminates sources with poor observational quality, unusual spectral shapes, and those exhibiting a double-peaked structure.

(iii) Additionally, we exclude all sources with peak frequencies $\nu_{\rm p} < 72$ \MHz 
to eliminate sources with few measurements of the optically thick spectrum or indistinct peak structures. Similarly, some sources with larger $\nu_{\rm p}$ exhibit the same issue, lacking sufficient measurements in the optically thin regime. However, since these sources are few in number, they are still included in our released catalog 
for possible future investigation with additional observations.

Finally, we identified 4,315 PS source candidates. Most of these sources are classified as MPS sources, with a small portion classified as GPS sources. 

\begin{figure}
\centering
\includegraphics[width=0.99\columnwidth]{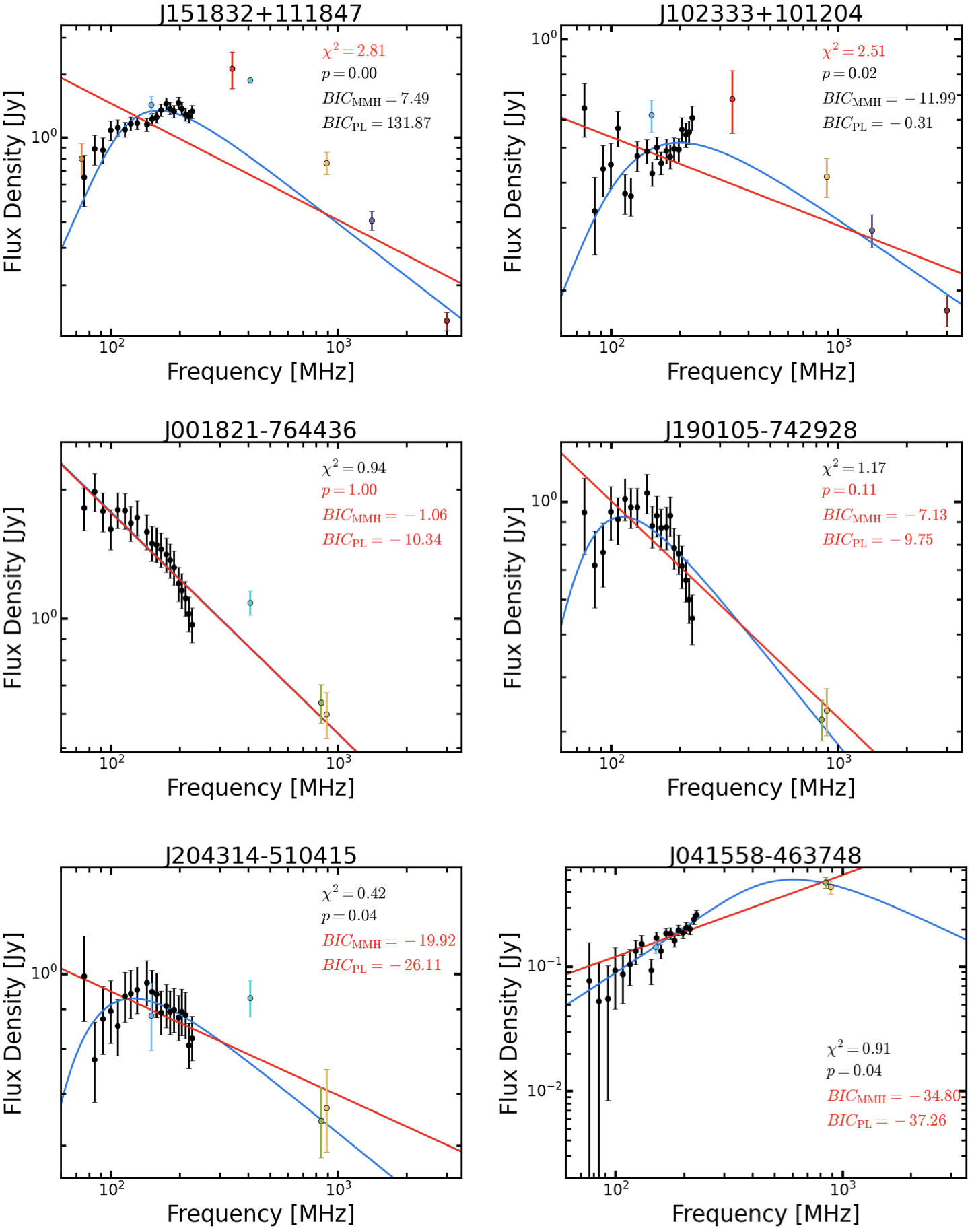}
\caption{Examples of sources excluded based on the p-value, BIC method, and $\chi^2_{\rm MMH}$ selection criteria. The blue solid line represents the fit of the MMH model, while the red solid line represents the power-law (PL) model. Error bars in different colors indicate different observational projects. In the right corner, the labels from top to bottom show the $\chi^2$ calculated based on the MMH Model, p-value, and the BIC values calculated for the two models, respectively. Factors leading to the exclusion of sources are highlighted in red.}
\label{fig:remove}
\end{figure}

Figure~\ref{fig:remove} shows examples of excluded radio sources based on the p-value, the BIC method, and the selection criteria $\chi^2_{\rm MMH}$, respectively. 
These sources lack coherent distinct peak features and are thus excluded by our selection criteria.
Additionally, during the cross-validation phase, only \textit{isolated} sources were accepted, leading to the removal of some previously identified double or multi-peaked sources. This suggests that the multi-peaked structures may be caused by multiple unresolved sources rather than being formed by a single source.

In our sample selection, we select only \textit{isolated} sources, but we do not exclude unresolved sources as in \cite{Callingham2017}, which avoided the possible contamination of multiple sources at the price of rejecting many genuine PS sources. 
Nevertheless, we provide the calculated extent factor $ab/(a_{\rm psf}b_{\rm psf})$ for reference, where $a$, $b$, $a_{\rm psf}$, and $b_{\rm psf}$ are the semimajor and semiminor axes of a source and the point-spread function, respectively \citep{Hurley-Walker2017}. Of the total PS source samples, 650 sources have $ab/(a_{\rm psf}b_{\rm psf})>1.1$.

\section{Results}
\label{sec:results}
The advantage of an MMH model is that it not only accurately fits the spectral shape of the PS but also provides meaningful physical insights. The model parameters allow for further investigation into the environmental structure surrounding PS sources. For example, it allows us to determine whether a source is primarily dominated by FFA absorption or SSA absorption through Bayesian fitting. Furthermore, as described in the formulas in Section~\ref{sec:A Hybrid Spectral Model}, the spectral analysis can reveal information about the magnetic field ($B$), electron density ($n_{\rm e}$), source scale, redshift, and other parameters around the radio source. This is crucial for understanding the formation and evolution of galaxies. In this section, we first compare our new model with previous studies, and then analyze the physical characteristics of PS sources based on the model fitting parameters.

\subsection{Comparison with previous models}
\label{sec: Compared to the previous model}
To evaluate the performance of the model, we compare our newly developed MMH model with the two best-performing models in \cite{Callingham2015}, 
i.e. the ``Inhomogeneous FFA” and ``Double SSA”. Following \cite{Callingham2015}, we also considered the effect of the spectral break
\citep[e.g.][]{Turner2018, Quici2021} on the fitting process by incorporating
an additional multiplicative factor in the model:
\begin{equation}
    F_{\rm br} = e^{(-\nu / \nu_{\rm br})}, 
	\label{eq:break}
\end{equation}
where $\nu_{\rm br}$ represents the high-frequency cutoff value due to spectral aging. 
Models that account for the spectral break are labeled as ``break” to distinguish them 
from those that do not. The spectral break term has little impact on the fitting 
results of other free parameters and generally affects higher frequency spectra more 
significantly \citep{Brienza2020, An2024}. 

\begin{figure*}
\centering
\includegraphics[width=\textwidth]{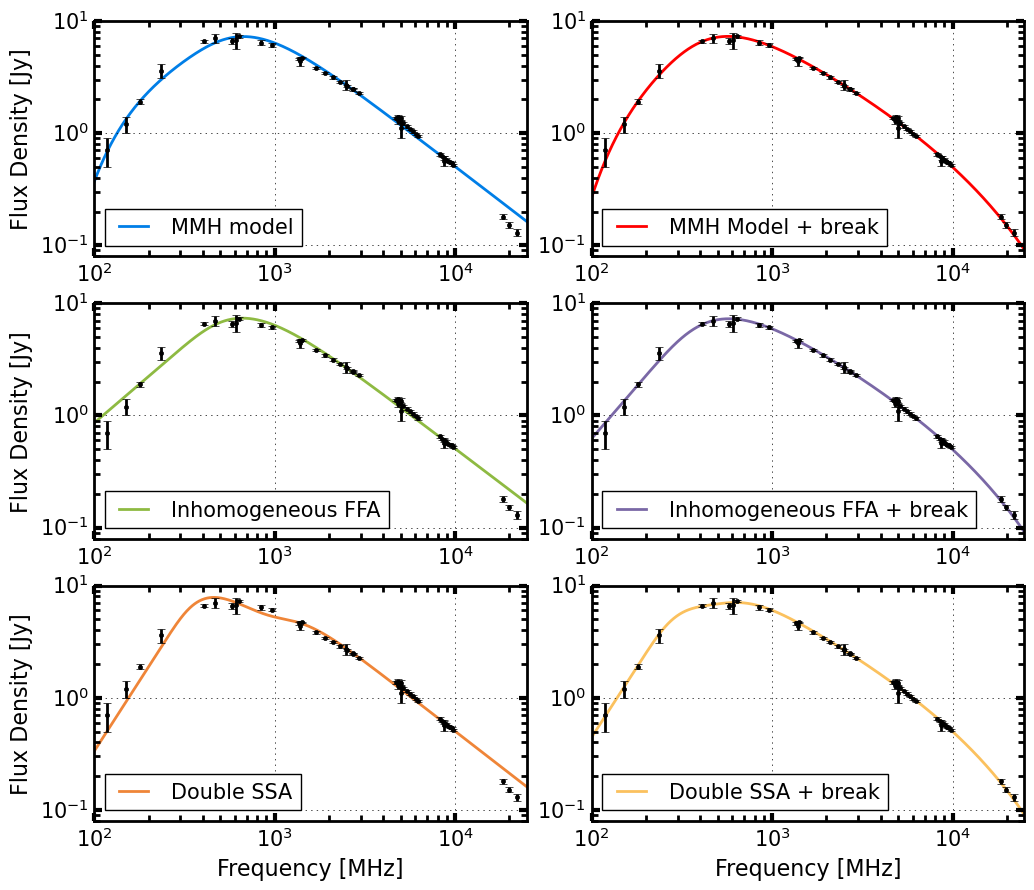}
\caption{The spectrum of PKS B0008-421 fitted by different models. 
The top panels show the ``MMH model" used in this paper, 
the middle panels show the model ``Inhomogeneous FFA”, and 
the bottom panels show the model ``Double SSA”. 
The right column panels show the fitting with the effect of spectral break. 
The ``MMH model" improves the goodness of fit at the low-frequency end and the
additional ``break'' mechanism improves fitting at the high-frequency end.}
\label{fig:compare}
\end{figure*}

We plot the spectrum measurements of PKS B0008-421 in Figure~\ref{fig:compare},
as well as the best-fit models in different panels. 
The models with the spectra break effect are shown in the right panels, and those without such an impact are shown in the left panels. 
In the low-frequency regime (72–500 \MHz), the MMH model provides significantly better fits compared to the other models. This advantage persists even when spectral break effects are excluded.

Table~\ref{tab:compare} presents the goodness-of-fit metrics ($\chi^2/$\text{d.o.f.} and BIC). Without considering spectral breaks, the MMH model achieves a $\chi^2/$\text{d.o.f.} = 3.0 for low-frequency data, outperforming the Inhomogeneous FFA model ($\chi^2/$\text{d.o.f.} = 5.5) and the Double SSA ($\chi^2/$\text{d.o.f.} = 8.1). When incorporating spectral breaks, the MMH model achieves a best-fit $\chi^2$/\text{d.o.f.} = 0.5, demonstrating superior accuracy in high-frequency regimes.

Unlike prior models, the MMH model allows detailed exploration of absorption mechanisms, enabling us to disentangle the contributions of SSA and FFA to the observed spectrum. It also facilitates the estimation of physical parameters such as magnetic field strength and electron density.

\begin{table*}
	\centering
	\caption{The goodness of fit as characterized by $\rm \chi^2/d.o.f$ and BIC for different models of the radio source PKS B0008-421. The subscript ``mid" indicates that the three high-frequency points were removed, while ``low" indicates that only the low-frequency points below 500 \MHz were considered.}
	\label{tab:compare}
	\begin{tabular}{lcccccccc}
		\hline
		Models & d.o.f & $\rm \chi^2/d.o.f$ & $\rm \chi^2_{\rm mid}/d.o.f$ & $\rm \chi^2_{\rm low}/d.o.f$ & BIC & $\text{BIC}_{\rm mid}$ & $\text{BIC}_{\rm low}$ \\
		\hline
        MMH Model & 5 & 5.8 & 3.0 & 3.0 & 66 & 6 & 15\\
        MMH Model + break & 6 & 0.5 & 0.5 & 0.2 & -54 & -44 & 7\\
        Inhomogeneous FFA & 4 & 6.2 & 3.5 & 5.5 & 73 & 13 & 21\\
        Inhomogeneous FFA + break & 5 & 0.5 & 0.5 & 0.3 & -58 & -48 & 7\\
        Double SSA & 6 & 5.9 & 3.7 & 8.1 & 73 & 26 & 33\\
        Double SSA + break & 7 & 0.5 & 0.5 & 0.1 & -50 & -40 & 10\\
        \hline
	\end{tabular}
\end{table*}

\begin{figure*}
\centering
\includegraphics[width=0.9\textwidth]{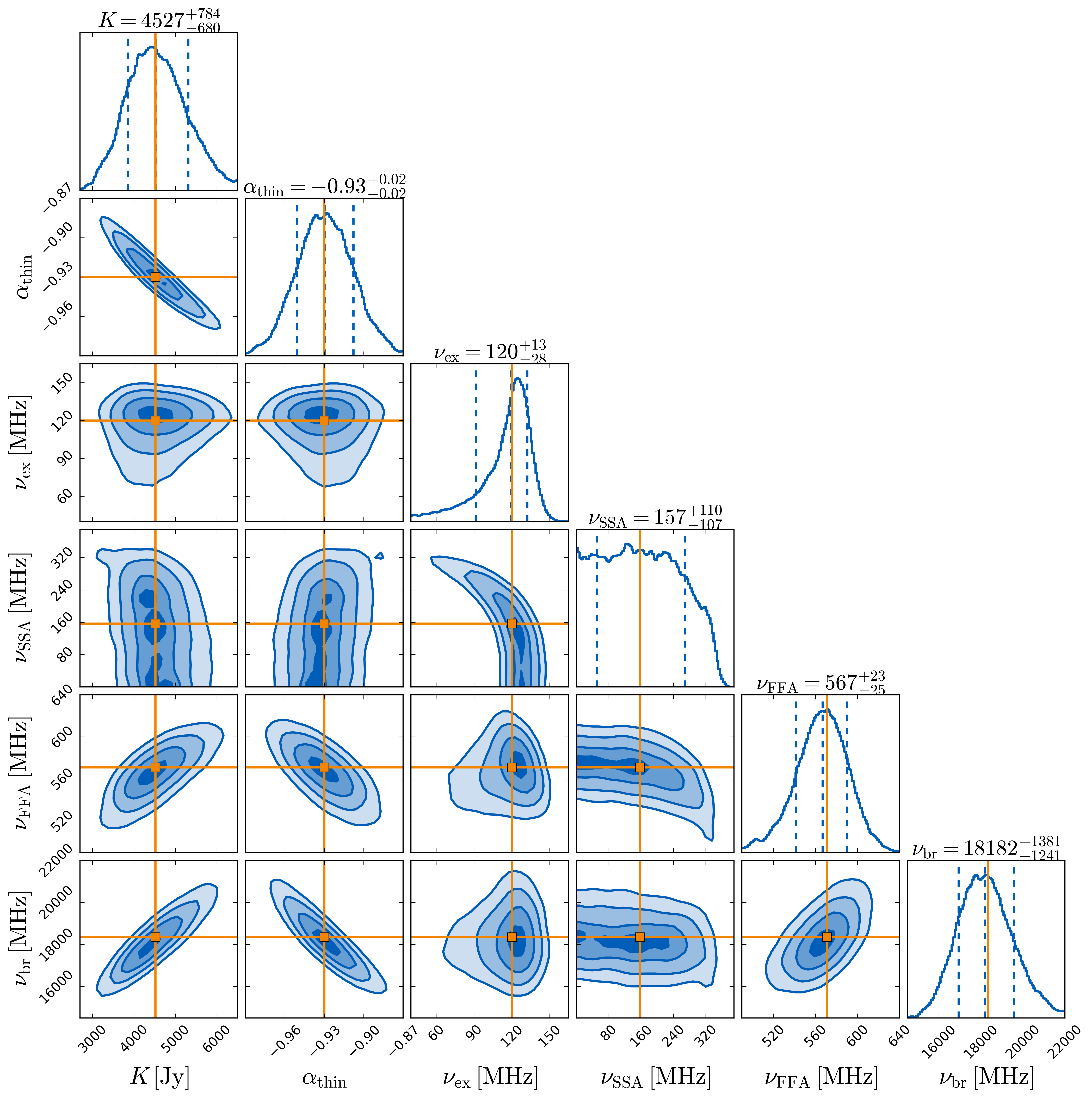}
\caption{Corner plot of the ``MMH Model + break" MCMC fit for PKS B0008-421. The blue dashed lines show the median of the distribution as well as the 16th and 84th percentiles. The solid orange line represents the parameters corresponding to the maximum likelihood. For some distributions, such as $\nu_{\rm FFA}$, which are highly non-Gaussian, we actually used the maximum likelihood instead of the median as the best-fit result. These parameters are listed in Table~\ref{tab:pksb0008}.}
\label{fig:mcmc}
\end{figure*}

\begin{table*}
	\centering
	\caption{The fitting parameters of PKS B0008-421 spectrum, and the flux density and frequency at the peak.}
	\label{tab:pksb0008}
	\begin{tabular}{lcccccccc}
		\hline
		Model & K & $\alpha_{\rm thin}$ & $\nu_{\rm ex}$ & $\nu_{\rm SSA}$ & $\nu_{\rm FFA}$ & $\nu_{\rm br}$ & $S_{\rm p}$ & $\nu_{\rm p}$\\
		\hline
        &(\Jy) & & (\MHz) & (\MHz) & (\MHz) & (\MHz) & (\Jy) & (\MHz)\\
        \hline
        MMH Model & 41813$\pm$2577 & -1.23$\pm$0.01 & 123$\pm$8 & 96$\pm$66 & 794$\pm$17 & --  & 7.26$\pm$0.11 & 665$\pm$10\\
        MMH Model + break & 4519$\pm$734 & -0.93$\pm$0.02 & 121$\pm$28  & 157$\pm$96 & 571$\pm$26 & 18344$\pm$1325 & 7.25$\pm$0.11 & 575$\pm$12 \\
        \hline
	\end{tabular}
\end{table*}

Figure~\ref{fig:mcmc} presents the 2-dimensional posterior distribution of the
fitting parameters for the ``MMH Model + break" case for PKS B0008-421. 
The median values and the maximum likelihood of the distribution are marked in the plots.
Table~\ref{tab:pksb0008} lists the values corresponding to the maximum likelihood value. 
The values of $\nu_{\rm FFA}$ and $\nu_{\rm p}$ are very close, indicating that the internal FFA is the main absorption mechanism that forms the peaked spectrum of this radio source.  

A flat posterior distribution of $\nu_{\rm SSA}$ suggests that the constraints on SSA 
in the observed frequency bands are weak, necessitating lower-frequency observations
for better constraints. 
The posterior probability distribution of $\nu_{\rm SSA}$ remains nearly uniform across the wide frequency range of $0$--$240\, \MHz$, suggesting a loose upper limit. To test this, we set $\tau_{\rm SSA} = 0$, and found that the spectral data could still be effectively fitted using only internal and external FFA mechanisms. This result further supports the conclusion that FFA is the primary absorption mechanism for PKS B0008-421.

Substituting the median value of the posterior probability distribution of $\nu_{\rm SSA}$,
as well as the observational properties of PKS B0008-421, i.e., its angular size
$\theta = 120\ {\rm mas}$ and $z = 1.12$ \citep{Callingham2015} into Equation~(\ref{eq:B}), 
we estimate the magnetic field strength to be in the range $0<B<0.1\ {\rm Gauss}$, with a median probable value of $B \approx 4.55\times 10^{-4}\ {\rm Gauss}$,
which is within a normal parameter range \citep{Orienti2008}.

Estimation of the mean electron density requires a detailed configuration of the AGN structure.
The physical size of the internal region of the AGN where the FFA mechanism dominates is assumed to be
$s^{\prime} \approx 1\ \kpc$, while the external region is assumed to be
$s \approx 20\ \kpc$ \citep{Callingham2015}.
The mean electron temperature has a negligible variation between the internal and external regions of the AGN \citep{Zhu2019, Jin2023, Hall2024}. We assume a single-electron temperature
$T_{\rm e} = T_{\rm e^{\prime}} \approx$ 10,000 K \citep{Reynolds1990, Haffner2009, Cong2021, Cong2022}. The electron number density may span several orders of magnitude.
According to the best-fit model parameters of $\tau_{\rm FFA}$ and $\tau_{\rm ex}$, 
the electron density is estimated as $n_{\rm e} \approx 66\ \rm cm^{-3}$ and 
$n_{\rm e} \approx 3\ \rm cm^{-3}$ for the internal and external region, respectively.
These values align well with the theoretical models of AGN \citep{Xiao2018, Zhu2019, Jin2023, Hall2024}, indicating that our model can explain the observed PS without invoking an extreme physical environment

\subsection{Statistical Properties of PS Sources}
\label{sec:Statistical Laws of PS Sources}

\begin{table}
	\centering
	\caption{The median value of the fitting uncertainties of the PS model. 
    Column ``$X$'' represents the model parameters, while
    the column of ``$\sigma_X$'' and  ``$\sigma_X/X$'' represent the absolute and
    the relative uncertainty of the corresponding parameter, respectively.
 }
	\label{tab:sigma}
\begin{tabular}{llll}
\hline
$X                 $ &        & $\sigma_X$      & $\sigma_X/X$ \\ \hline
$K                 $ & (\Jy)  & $23  $& $0.44$ \\
$\alpha_{\rm thin} $ &        & $0.06$& $0.07$ \\
$\alpha_{\rm thick}$ &        & $2.50$& $0.93$ \\
$\nu_{\rm ex}      $ & (\MHz) & $22  $& $1.38$ \\
$\nu_{\rm SSA}     $ & (\MHz) & $26  $& $1.00$ \\
$\nu_{\rm FFA}     $ & (\MHz) & $36  $& $0.35$ \\
$S_{\rm p}         $ & (\Jy)  & $0.02$& $0.05$ \\
$\nu_{\rm p}       $ & (\MHz) & $11  $& $0.10$ \\ 
\hline
\end{tabular}
\end{table}

We introduce the spectral index $\alpha_{\rm thick}$ in the low-frequency-end of the peaked spectrum, 
i.e. $\nu \ll \nu_{p}$, where it is optically thick, i.e. $\tau \gg 1$. 
The value of $\alpha_{\rm thick}$ is obtained by fitting the spectral data in the range $0.1$--$0.4 \nu_{\rm max}$ with an empirical power-law model, where $\nu_{\rm max}$ is the maximum value among $\nu_{\rm ex}$, $\nu_{\rm FFA}$, and $\nu_{\rm SSA}$, representing the absorption mechanism that has the greatest influence on the peak.
The median values of the fitting uncertainties of $\alpha_{\rm thick}$, as well as the MMH parameters, are given in Table~\ref{tab:sigma}. 

\begin{figure}
\centering
\includegraphics[width=\columnwidth]{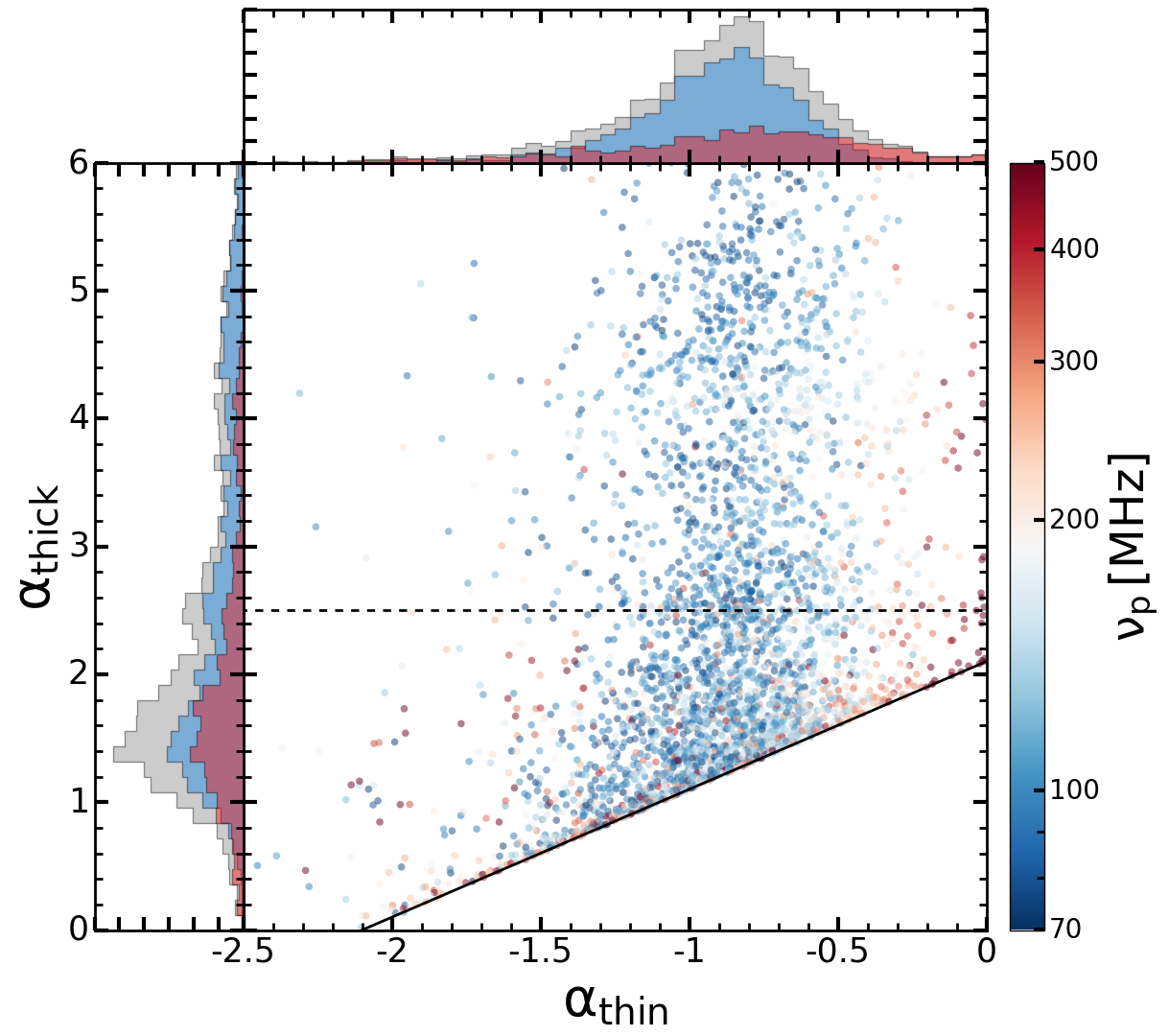}
\caption{The radio spectrum $\alpha_{\rm thick}-\alpha_{\rm thin}$  diagram of PS sources. 
Each point represents a PS source. Different colors correspond to different peak frequencies 
fitted with our MMH model. 
The solid black line depicts the $\alpha_{\rm thick,\,FFA}$ (i.e. Equation~(\ref{eq:thickFFA}))
and the dashed black line indicates the $\alpha_{\rm thick,\,SSA}$ (i.e. Equation~(\ref{eq:thickSSA})).
The histograms along the horizontal and vertical axes show the statistical distribution of 
$\alpha_{\rm thin}$ and $\alpha_{\rm thick}$, respectively. 
The red histograms indicate the distribution with the sources' peak frequency 
$\nu_{\rm p}\geq150$\MHz, the blue ones show results with $\nu_{\rm p}<150$\MHz, 
and the combinations of all sources are shown with the gray histograms. 
}
\label{fig:thickthin}
\end{figure}

\begin{figure*}
\centering
\includegraphics[width=\textwidth]{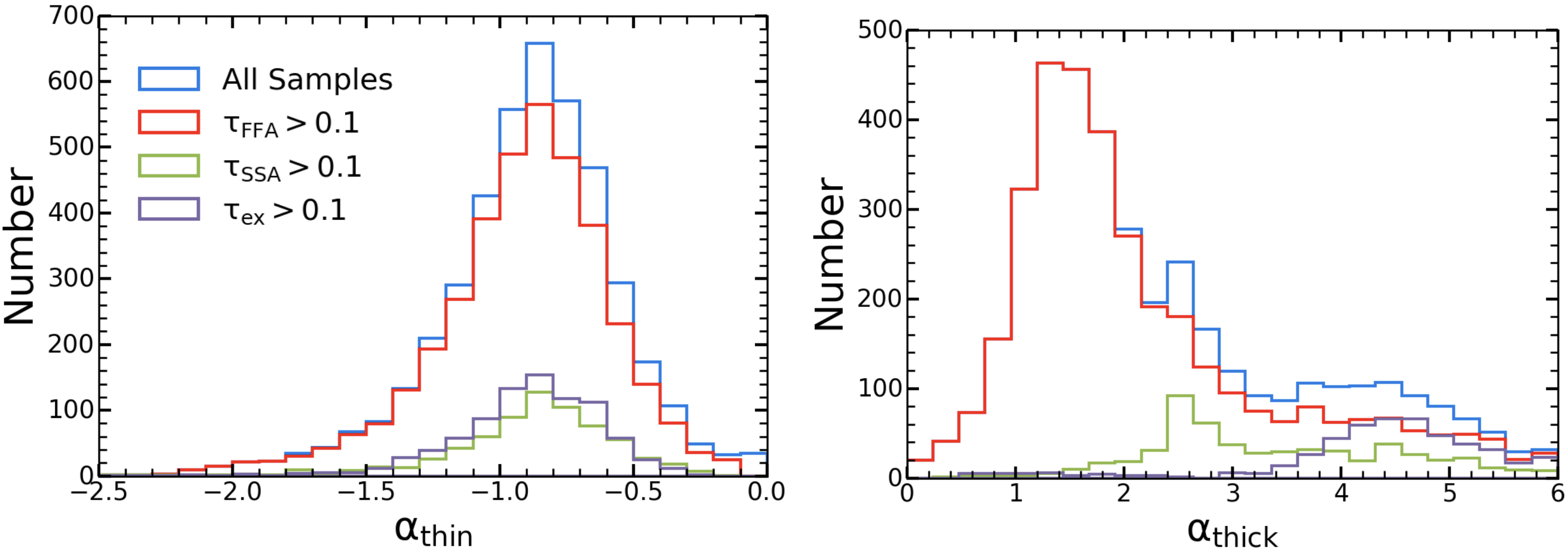}
\caption{The left and right panels show the histogram distribution of the 
fitted $\alpha_{\rm thick}$ and $\alpha_{\rm thin}$. 
Different colors represent the major absorption mechanism, which has the optical depth at the
peak frequency greater than $0.1$, i.e. the red is FFA absorption dominated, 
the green is internal SSA dominated, and purple is external FFA dominated, respectively.
Note that the same source may have multiple absorption optical depths greater than 0.1. 
}
\label{fig:thinthick}
\end{figure*}

\begin{figure}
\centering
\includegraphics[width=\columnwidth]{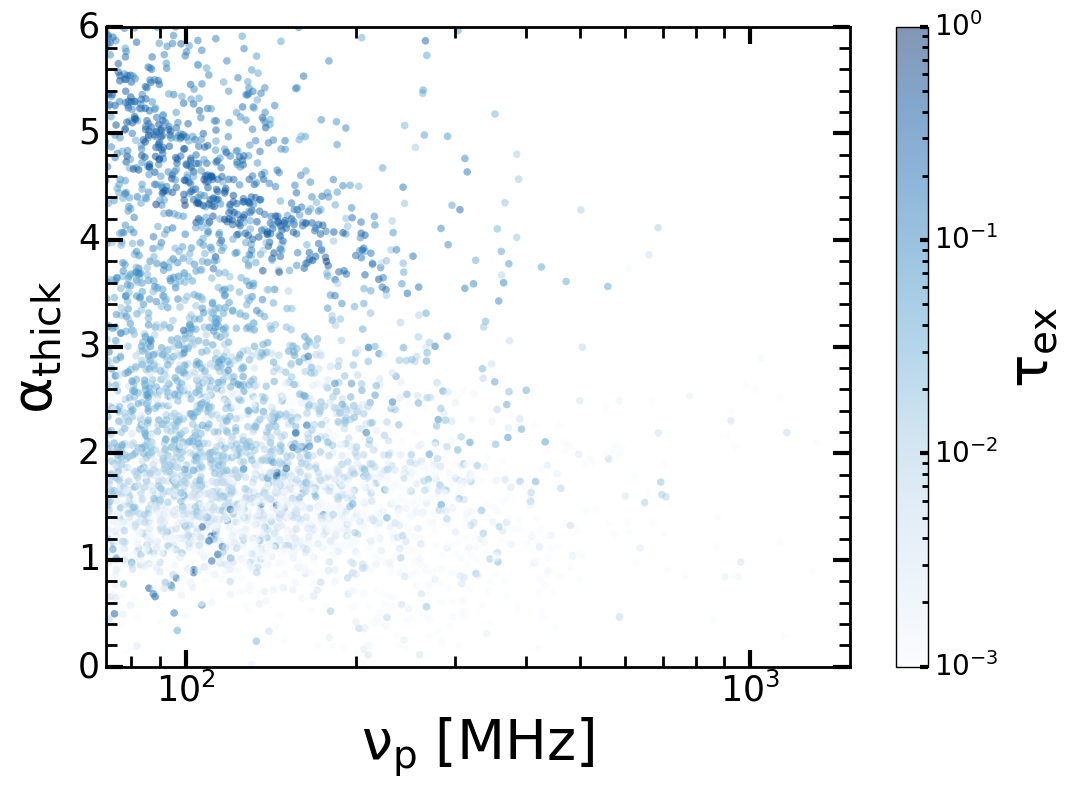}
\caption{The relationship between the observed peak frequency and the spectral index in the optically thick region. With the color of the points representing the magnitude of $\tau_{\rm ex}$ at $\nu_{\rm p}$. The darker the color, the greater the impact of ``external" absorption on the peak. 
The distribution of $\alpha_{\rm thick}$ follows two distinct trends. One remains around 1.5, almost unaffected by $\nu_{\rm p}$, while the other increases as $\nu_{\rm p}$ decreases, with this group of sources primarily dominated by $\tau_{\rm ex}$.}\label{fig:vpthic}
\end{figure}

At the lower frequency end,  i.e. $\nu\ll \nu_{\rm SSA}$ and  $\nu \ll \nu_{\rm FFA}$,
the exponential term in Equation~(\ref{eq:pspmodel}) associated with SSA and internal FFA vanished quickly, and Equation~(\ref{eq:pspmodel}) results in an optically thick spectrum,
\begin{equation}\label{eq:thick}
\left(\frac{\nu}{\rm MHz}\right)^{\alpha_{\rm thick}} = 
\frac{K\big(\nu/{\rm MHz}\big)^{\alpha_{\rm thin}}\times e^{ -(\nu/\nu_{\rm ex})^{-2.1} }} {\big(\nu/\nu_{\rm SSA}\big)^{\alpha_{\rm thin} - 2.5} + \big(\nu/\nu_{\rm FFA}\big)^{-2.1}},
\end{equation}
where the remaining exponential term is associated with the external FFA absorption. 
In a simplified case, only internal FFA absorption exists and both the SSA and external FFA are negligible, we then have a relationship between $\alpha_{\rm thick}$ and $\alpha_{\rm thin}$,
\begin{equation}\label{eq:thickFFA}
{\alpha_{\rm thick,\,FFA}} = {\alpha_{\rm thin} + 2.1}.
\
\end{equation}
On the other hand, if the SSA dominates the absorption, it theoretically results in a constant $\alpha_{\rm thick}$ expressed as,
\begin{equation}\label{eq:thickSSA}
{\alpha_{\rm thick,\,SSA}} = 2.5.
\end{equation}
With the combination of different absorption mechanisms, the spectrum becomes steeper, 
and $\alpha_{\rm thick}$ becomes larger.
Considering the exponential term of the external FFA absorption,
$\alpha_{\rm thick}$ increases as the frequency decreases, resulting in a spectrum that deviates from a pure power-law.

Figure~\ref{fig:thickthin} shows the fitted $\alpha_{\rm thin}$ and $\alpha_{\rm thick}$ for all PS sources in our sample. The color indicates the peak frequency $\nu_{\rm p}$ in the MMH model.
The solid black line marks $\alpha_{\rm thick,\,FFA}$ given in Equation~(\ref{eq:thickFFA}), and the black dashed line shows $\alpha_{\rm thick,\,SSA}$ given in Equation~(\ref{eq:thickSSA}).
$\alpha_{\rm thick,\,FFA}$ gives a lower bound of $\alpha_{\rm thick}$.
Any deeper frequency spectrum indicates the existence of more complex absorption mechanisms.

The histograms along the horizontal and vertical axes of Figure~\ref{fig:thickthin}
show the statistical distribution of $\alpha_{\rm thin}$ and $\alpha_{\rm thick}$.
It shows that sources with smaller peak frequency $\nu_{\rm p}$ have a more dispersed distribution of $\alpha_{\rm thick}$ and, 
comparing to the sources with larger $\nu_{\rm p}$, they slightly favor deeper spectrum index.
The distribution of $\alpha_{\rm thin}$ is relatively concentrated at $\alpha_{\rm thin} \sim -0.9$ for sources with larger $\nu_{\rm p}$.

For the pure theoretical SSA mechanism, the optically thick spectrum index is a constant 2.5.
We observe a minor peak in the $\alpha_{\rm thick}$ histogram near the black dashed line in Figure~\ref{fig:thickthin}, indicating that some PS sources are probably dominated by SSA.
Because the characteristic frequency of external FFA, $\nu_{\rm ex}$, can extend to low frequencies, it also affects the optically thick spectral index and contributes to the broad scatter in the $\alpha_{\rm thick} - \alpha_{\rm thin}$ plane.

Figure~\ref{fig:thinthick} shows the statistical histograms for $\alpha_{\rm thin}$ and
$\alpha_{\rm thick}$. Here we also plot the sub-samples according to their major absorption mechanism, which are defined as having an absorption optical depth of that mechanism greater than 0.1 at peak frequency.
However, there could be multiple absorption mechanisms that contribute significantly, so these subsamples do have some overlap.

Figure~\ref{fig:thinthick} shows that sources dominated by different absorption mechanisms have almost identical $\alpha_{\rm thin}$ distributions.
However, their $\alpha_{\rm thick}$ distributions differ. Most PS sources in our sample are dominated by FFA; for those with $0 < \alpha_{\rm thick} < 2$, internal FFA alone seems sufficient to explain the absorption. For $2 < \alpha_{\rm thick} < 3.5$, SSA also plays an important role alongside internal FFA. Meanwhile, when $\alpha_{\rm thick} > 3.5$, external FFA often becomes a major mechanism as well, in addition to internal FFA and SSA.

Figure~\ref{fig:vpthic} plots the fitted parameters in the $\alpha_{\rm thick}$--$\nu_{\rm p}$ space, with each point color-coded by its $\tau_{\rm ex}$. Sources dominated by external FFA generally exhibit larger $\alpha_{\rm thick}$ values, which show a weak correlation with the peak frequency—lower $\nu_{\rm p}$ corresponds to higher $\alpha_{\rm thick}$. However, $\alpha_{\rm thick}$ from model fitting can have large uncertainties, especially in sources with lower $\nu_{\rm p}$, where limited low-frequency coverage constrains the fit. High-sensitivity observations at even lower frequencies are needed to better quantify this relationship.

\begin{figure*}
\centering
\includegraphics[width=\textwidth]{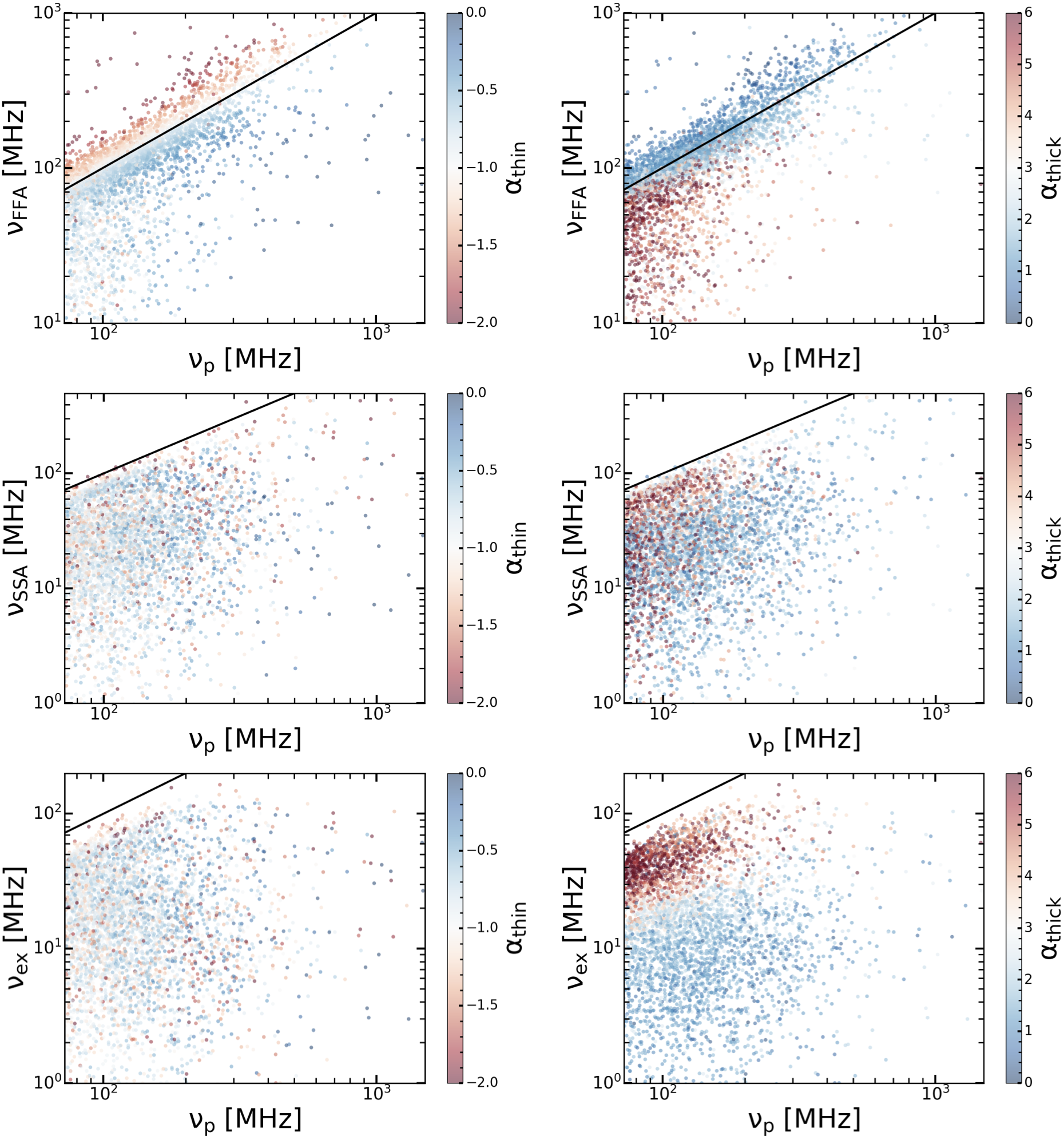}
\caption{Relationship diagram between the characteristic frequency of different absorption 
mechanisms and the peak frequency $\nu_{\rm p}$. 
The black solid line indicates where the characteristic frequency equals $\nu_{\rm p}$. 
The color of the left panel represents the optically thin spectral index $\alpha_{\rm thin}$, 
while the color of the right panel represents the optically thick spectral index $\alpha_{\rm thick}$. 
The y-axis from top to bottom represents the characteristic frequency of internal FFA, SSA, 
and external FFA absorption, respectively.}
\label{fig:vp}
\end{figure*}

Figure~\ref{fig:vp} shows the relationship between the characteristic frequencies of 
different absorption mechanisms and the peak frequency $\nu_{\rm p}$. 
When the characteristic frequency is close to $\nu_{\rm p}$, the corresponding absorption mechanism produces significant optical depth and thus dominates the low-frequency spectrum. 
The solid black line indicates the condition where the characteristic frequency equals $\nu_{\rm p}$, in which case the optical depth at the peak frequency is unity.
We observe that a large fraction of PS sources have $\nu_{\rm FFA}$ constrained near $\nu_{\rm p}$, implying that internal FFA is a major absorption mechanism in our PS sample. 
PS sources dominated by internal FFA typically exhibit small $\alpha_{\rm thick}$.
On the other hand, PS sources with large $\alpha_{\rm thick}$ generally have 
$\nu_{\rm ex}$ close to $\nu_{\rm p}$. These findings are consistent with the statistical histograms shown in Figure~\ref{fig:thinthick}.

\section{Discussion}
\label{sec: Discussion}

\subsection{Flat Low-Frequency Spectrum}
\label{sec: Flat Low-Frequency Spectrum}

In the preceding section, we introduced sources dominated by SSA, for which the theoretical optically thick spectral index $\alpha_{\rm thick}$ is a constant value of 2.5. However, it is important to emphasize that various mechanisms can cause the spectrum to flatten at frequencies lower than the peak, resulting in an actual $\alpha_{\rm thick}$ value less than 2.5 for SSA-dominated sources. \citet{Cotton1980} proposed that the superposition of multiple components with different peak frequencies can produce an almost flat spectrum. \citet{Condon2016} pointed out that real astrophysical sources are inhomogeneous, and the combination of unresolved components can lead to a spectral index below 2.5. Such multiple components may originate from jets (CSS or CSO) \citep[e.g.][]{1979ApJ...232...34B}, inherent inhomogeneity of the galaxy’s magnetic field \citep[e.g.][]{Artyukh2008, 2019ApJ...873...55B, 2020ApJ...894...47B}, or two galaxies closely aligned along the line of sight.

With limitations in observational resolution, this issue will persist. For example, in the six PS sources in \cite{Keim2019} where high-resolution VLBA observations are available, three of these sources exhibit multi-component structures. However, in two of these sources, the different components show significant differences in either brightness or spectral index. In such cases, i.e., one component is dominant while the other is faint, the influence of the faint component is small and negligible.

Furthermore, although CSS or CSO sources may exhibit asymmetries in lobe intensity and turnover frequency, as long as multiple components possess similar physical structures and absorption mechanisms, the MMH model remains sufficient to explain them under low-resolution requirements.
The contribution fraction of different absorption mechanisms can be determined statistically by fitting with the observation data.
Therefore, we emphasize that it is more likely the different absorption mechanisms are responsible for the PS source than the multiple components.

In our hybrid model, the source spectrum is shaped by the combined effects of internal FFA, SSA, and external FFA mechanisms, with any of these potentially dominating. However, as shown in Section~\ref{sec:results} and illustrated in Figure~\ref{fig:thinthick},
approximately $50\%$ of the PS sources identified in the current catalog have $\alpha_{\rm thick} < 2$, with the distribution peaking around $\alpha_{\rm thick} \sim 1.5$. 

Addressing the initial question, we tested whether the assumption of $\alpha_{\rm thick}=2.5$ prevents SSA from being the dominant absorption mechanism. Specifically, considering the inhomogeneity of the source, we incorporated $\alpha_{\rm thick}$ as a free parameter into Equation~(\ref{eq:tau_ssa}) to investigate the impact of varying $\alpha_{\rm thick}$ on the model constraints.
\begin{equation}
    \tau_{\rm SSA} = \left( \frac{\nu}{\nu_{\rm SSA}} \right) ^{\alpha_{\rm thin} - \alpha_{\rm thick,\, SSA}}.\label{eq:tau_ssa_modify}
\end{equation}
The results indicate that the proportion of PS sources dominated by SSA has significantly increased; however, it still does not surpass the number of sources dominated by internal FFA. Moreover, this adjustment leads to a large number of sources with $\alpha_{\rm thick,\,SSA}$ values approaching zero. Sources exhibiting such low values are more likely formed by the superposition of multiple SSA components. Additionally, some sources with $\alpha_{\rm thick}$ slightly below 2.5 may be influenced by inhomogeneities in the magnetic field. The specific effects of these magnetic field inhomogeneities on $\alpha_{\rm thick,\,SSA}$ require further observational
 physical explanation \citep{2019ApJ...873...55B, 2020ApJ...894...47B}, which we reserve for future work.

Flat optically thick spectra can also be directly achieved through the internal FFA mechanism. As shown in Equation~(\ref{eq:thickFFA}), internal FFA sets the lower limit of $\alpha_{\rm thick}$, which is a function of $\alpha_{\rm thin}$. For sources entirely dominated by internal FFA, there is a strong correlation between the spectral indices in the optically thick and optically thin regions, as depicted by the black solid line in Figure~\ref{fig:thickthin}. Many sources closely follow this line, indicating that FFA sufficiently and straightforwardly explains these sources. Additionally, sources dominated by internal FFA should also experience spectral flattening due to source inhomogeneity (unresolved multiple components), which would result in some sources appearing below the black solid line. However, we did not find any such sources in Figure~\ref{fig:thickthin}, suggesting that the probability of this scenario is very low. Therefore, our analysis indicates that although the number of SSA-dominated sources may be higher than currently presented, internal FFA remains the primary absorption mechanism.

\subsection{The major absorption mechanism}
\label{sec: Why is FFA so important?}

\begin{figure*}
\centering
\includegraphics[width=\textwidth]{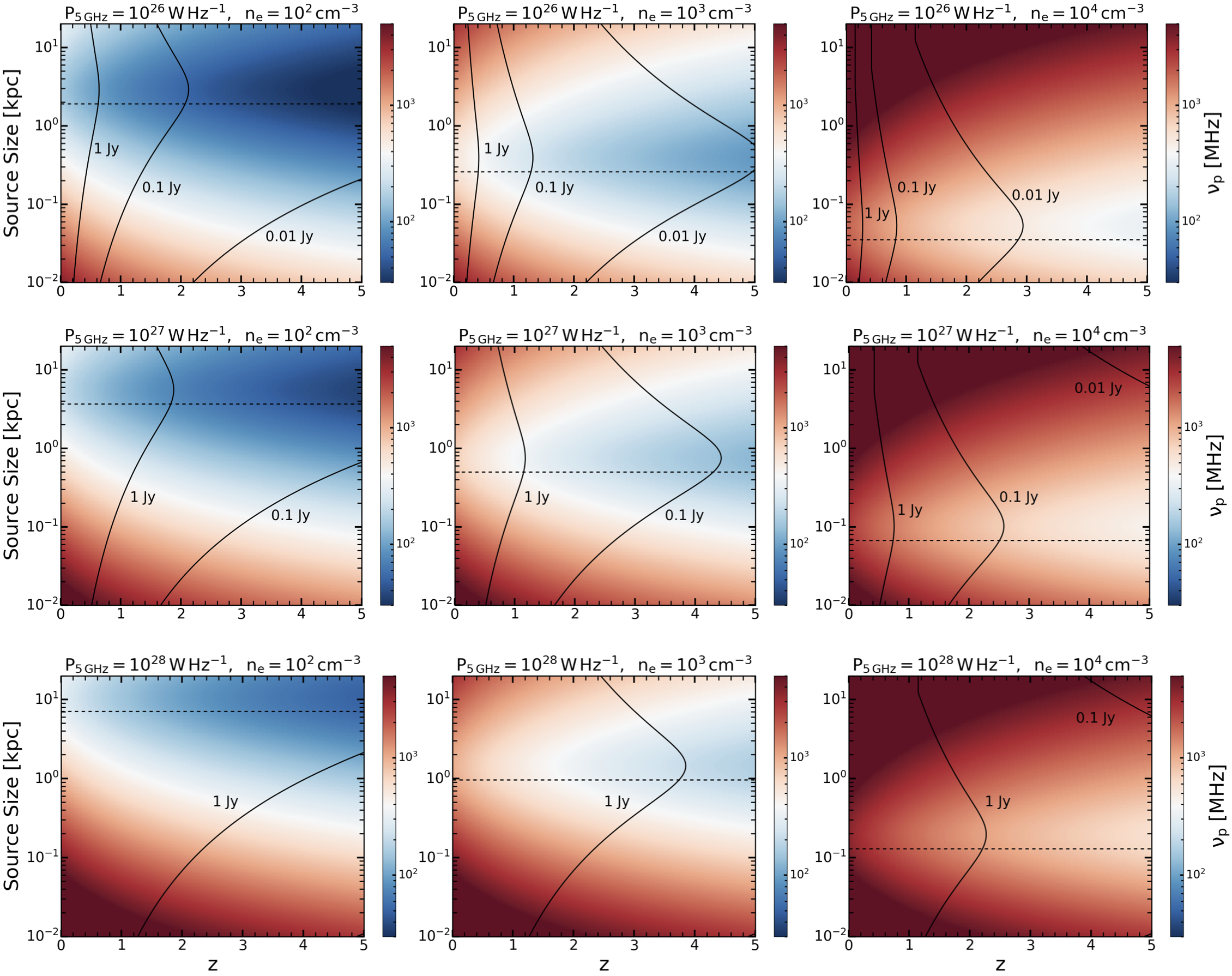}
\caption{The distribution of peak frequencies calculated over a broad parameter space. The horizontal dashed line represents the boundary between SSA and FFA, with the region above the line indicating FFA dominance and the region below indicating SSA dominance. $\nu_{\rm p}$ is the observed value, chosen as the larger value between the calculated  $\nu_{\rm SSA}$  and  $\nu_{\rm FFA}$. The black curve represents the observed flux density at the peak, with annotations nearby; the left side indicates higher brightness. The values of $P_{\rm 5\,GHz}$ and $n_{\rm e}$ used are noted at the top of each panel.}
\label{fig:P5ne}
\end{figure*}

In this section, we will explore the primary absorption mechanisms of PS sources based on the MMH model within the complete physical parameter space.
We omit external FFA because it is typically subdominant at commonly observed frequencies and mainly affects the very low-frequency end.
As shown in the right panel of Figure~\ref{fig:thinthick}, the external FFA is more effective at low frequencies and would result in a steeper spectrum with $\alpha_{\rm thick}\gtrsim3$. Such an absorption effect only dominates a small portion of the PS sources. As depicted in Figure~\ref{fig:thinthick}, less than 20\% of the sources exhibit a peak optical depth exceeding 0.1. Furthermore, Figure~\ref{fig:vp} illustrates that these sources typically deviate most significantly from the black solid line representing the dominant peak.

According to the physical characteristics of the PS sources, the initial set of physical parameters includes the source size $s^{\prime}$, radiation power $P_{5\,{\rm GHz}}$ at $5$ \GHz, 
magnetic field strength $B$, electron temperature and density, and the redshift of the source. 
According to Equation~(\ref{eq:nu_ssa}), the peak frequency is less sensitive to $B$.
We fix $B = 0.1\,{\rm Gauss}$\citep{Orienti2014, Keim2019}, which is high enough to avoid underestimating the influence of SSA. 
In addition, the parameter space for $T_{\rm e}$ is typically small. 
We fix $T_{\rm e} = 10,000\,{\rm K}$ and vary $n_{\rm e}$ between $100$, $1,000$ and $10,000\, {\rm cm}^{-3}$ \citep{Reynolds1990, Artyukh2008, Moe2009, Bicknell2018, Riffel2021}. 
$P_{5\,{\rm GHz}}$ relates to the observed flux density as: 

\begin{equation}
    \left(\frac{P_{5\,{\rm GHz}}}{\rm W\,Hz^{-1}}\right) = 4 \pi \left( \frac{D_{L}}{\rm m} \right)^2  
    \left( \frac{S_{5\,{\rm GHz}}}{\rm 10^{26}\ Jy} \right) (1+z)^{-(1+\alpha_{\rm thin})}, 
	\label{eq:P5GHz}
\end{equation}
where $(1+z)^{-(1+\alpha_{\rm thin})}$ is the k-correction, 
$\alpha_{\rm thin}$ is set as a constant value of $-0.7$, informed by previous studies and designed for broader applicability across various sources \citep[e.g.][]{ODea1998, Snellen2000, Callingham2017}. $S_{5\,{\rm GHz}}$ is the flux density of the source at 5 \GHz.

Figure~\ref{fig:P5ne} illustrates the peak frequencies of PS sources under different physical parameters. In each panel, the dashed black line marks the boundary separating $\nu_{\mathrm{SSA}}$ and $\nu_{\mathrm{FFA}}$ dominated regimes. Above the dashed line, the PS sources are dominated by internal FFA, whereas below it, SSA is the primary absorption mechanism.

In Figure~\ref{fig:P5ne}, as $n_{\rm e}$ increases from the left to right columns, the FFA is affected primarily, leading to an increase in $\nu_{\rm p}$ for the FFA dominated sources and pushing the dashed line downward. From the top to bottom panels, as the radiation power increases, the SSA is predominantly affected, resulting in an increase in $\nu_{\rm p}$ for the SSA-dominated sources and pushing the dashed line upward. Additionally, the dominant mechanism is independent of the redshift. The black solid lines in the figure indicate the flux density at the peak, with a higher flux density on the left side of the line. Assuming a uniform distribution of source sizes on a logarithmic scale, FFA and SSA each dominate approximately $50\%$ of the PS sources.

Figure~\ref{fig:P5ne} also shows that PS sources with peak frequencies around \GHz
(shown in red) are mostly associated with compact high-power SSA-dominated sources or FFA sources which are large in size. 
These two mechanisms differ significantly in linear scale and can be well distinguished with higher angular resolution observations ($<100$ mas). 
At lower frequencies, e.g. around a hundred \MHz, sources are larger in size and generally have higher redshifts. The current investigation assumes a high magnetic field strength. 
In reality, the magnetic field strength is probably lower than what we assumed, and the peak frequency could be even lower.
Therefore, future ultra-long wavelength observations of extragalactic radio sources 
are expected to discover more PS sources at high redshift.

However, the parameters are not fully independent. For instance, $n_{\rm e}$ and source size exhibit an inverse relationship \citep{Moe2009, Nicastro1999, Huerta2014}, so larger radio sources generally possess lower electron densities $n_{\rm e}$. Distinguishing the dominant absorption mechanism thus requires considering more complex parameter correlations, which rely on sufficient observational data. We defer a detailed analysis of these correlations to future work.

\subsection{Factors affecting radio spectra}
\label{Factors affecting radio spectra}

The MMH model only considers three mechanisms that could modify the power-law shape of radio spectra. 
Our results show that these three mechanisms have a significant impact on the spectra of bright AGN sources within the observed frequency range. 

Nonetheless, there are a couple of other possible scenarios or mechanisms that could also cause a non-power-law radio spectrum. 
For example, in Section~\ref{sec: Compared to the previous model}, we include the spectral breaks caused by spectral aging \citep{Turner2018, Quici2021} and inverse-Compton losses \citep{Potter2013, klein2018} to effectively explain the high-frequency spectral shape of source PKS B0008-421. 
However, spectral breaks typically occur at higher frequencies.
Models that account for these effects can significantly improve the spectral fit at the high-frequency end, but they are negligible for the low-frequency range targeted by the MMH model. 
Similarly, bremsstrahlung radiation mainly affects higher frequencies (\GHz) and faint galaxies \citep{klein2018}. Therefore, it is unlikely to significantly affect our sample.

The Razin–Tsytovich effect \citep{Melrose1980, Dougherty2003, ravi2019} arises because the refractive index of the plasma deviates from unity at low frequencies, reducing the efficiency of synchrotron emission by relativistic electrons. 
It could also cause additional absorption in the low-frequency spectrum, at

\begin{equation}
    \left( \frac{\nu}{\MHz} \right) \lesssim\ \left( \frac{\nu_{\rm R}}{\MHz} \right) \approx\ 20 \left( \frac{n_{\rm e}}{10^6\ \rm cm^{-3}} \right) \left( \frac{B}{\rm Gauss} \right)^{-1},
	\label{eq:rz}
\end{equation}
where $\nu_{\rm R}$ is the characteristic cut-off frequency of the Razin-Tsytovich effect. It introduces an additional multiplicative factor to the spectrum:
\begin{equation}
    F_{\rm R}=e^{-\nu_{\rm R}/\nu}.
	\label{eq:frz}
\end{equation}
For extreme parameters, such as $n_{\rm e} = 10^4\ {\rm cm^{-3}} $ and $B = 10^{-3}$ Gauss,  $\nu_{\rm R} = 200$ \MHz. 
However, the inclusion of this effect significantly increases the complexity of the model. 
Future studies, especially those that delve deeper into lower frequencies, are advised to consider this effect.

Ionization losses \citep{murphy2009, Basu2015, mckean2016} of the charged particles can alter the energy distribution of the particles that generate the radio radiation via the synchrotron process. The critical frequency $\nu_c$ given by classical synchrotron radiation theory is expressed as follows 
\begin{equation}
    \nu_c = \frac{3eB\gamma^2}{4\pi m_{\rm e}c},
	\label{eq:nuc}
\end{equation}
where $\gamma$ is the Lorentz factor of the electron, $e$ is the electron charge,  $m_e$ is the electron mass. The frequency it affects is \citep{1965ARA&A...3..297G, Condon2016}
\begin{equation}
    \left( \frac{\nu}{\MHz} \right) \approx 0.29\,\nu_c \approx 1.2 \, \gamma^2 \,\left( \frac{B}{\rm Gauss} \right).
	\label{eq:losses}
\end{equation}
Assuming a minimum Lorentz factor is $\gamma_{\min} = 100$, and $B = 0.001$ Gauss, the lower limit of the affected frequency is estimated to be around 12 \MHz. Therefore, ionization losses are unlikely to significantly affect sources within the observed range.

In summary, our statistical study aims to explain the turnover of PS sources as concisely as possible while providing physically meaningful fitting parameters, and identifying the primary causes that may influence spectral turnover. Therefore, mechanisms unlikely to strongly affect the spectral peak are not considered in our analysis. 
However, as observational frequencies decrease, these effects could become more pronounced, and the lower-frequency sky might prove darker than we currently anticipate.

\section{Conclusions}
\label{sec:Conclusions}

In this paper, we have developed a new multi-mechanism hybrid (MMH) model for Peaked-Spectrum (PS) sources, offering a realistic and physically meaningful framework for understanding these enigmatic objects. Employing a comprehensive dataset from various surveys, including GLEAM, VCSS, SUMSS, RACS, NVSS, and VLASS, we have identified and analyzed a significant sample of PS sources, enhancing our understanding of their physical properties and absorption mechanisms. The main results of this paper are as follows.

(i) The new model, inspired by the supernova parameterization, effectively combines synchrotron radiation emission with multiple absorption mechanisms, including synchrotron self-absorption (SSA) and free-free absorption (FFA). This model not only fits the observed spectra of PS sources accurately but also provides insights into the physical conditions within and around these sources.

(ii) Applying the MMH model, we identified a sample of $4,315$ PS sources. This identification method aims to comprehensively recognize sources with peak frequencies approximately between $72$--$3000$ \MHz. Most of these sources are classified as MHz-Peaked-Spectrum (MPS) sources, while a small portion can be classified as GHz-Peaked-Spectrum (GPS) sources.

(iii) Our analysis underscores the critical role of FFA in the spectral turnover of PS sources. Approximately $50\%$ of the sources exhibit $\alpha_{\rm thick} < 2$, suggesting that internal FFA is the dominant absorption mechanism shaping their spectral turnover (Although the possibility of SSA-dominated cannot be completely excluded.). This conclusion is further supported by the observed correlation between $\alpha_{\rm thick}$ and $\alpha_{\rm thin}$, which aligns with theoretical expectations for sources where thermal electron populations within the emitting region significantly absorb synchrotron radiation.

(iv) Sources with $\alpha_{\rm thick} > 3.5$ often display a substantial contribution from external FFA. This indicates that the extended external medium, likely consisting of dense ionized gas surrounding the AGN, plays a significant role in absorbing low-frequency emission. External FFA provides a compelling explanation for the steep spectral indices observed in these cases, particularly in the optically thick regime.

(v) By analyzing a large sample of PS sources, we identified several statistical properties that shed light on their nature. For instance, the relationship between the spectral indices of the optically thin and thick spectra reveals boundaries consistent with the SSA and FFA theories. Additionally, sources with lower peak frequencies tend to exhibit higher and more dispersed spectral indices in the optically thick region.

As noted in the Introduction, the absorption mechanism is related to the interpretation of the nature of the source, e.g. the ``youth'' hypothesis, in which the SSA is the primary absorption mechanism; and the ``frustration'' hypothesis, where the FFA dominates \citep[][]{Callingham2015, Bicknell2018, Keim2019, ODea2021}. 
So our results which favor the FFA as the primary absorption mechanism for the majority of the sources would also provide an important clue to the solution to the problem, though of course the hypotheses also need to be examined from many different perspectives, especially the consistency of the physical parameters obtained from different kind of observations. 

Future low-frequency radio surveys, such as those conducted by MWA, LOFAR, and SKA, will play an important role in advancing our understanding of PS sources. These observations will provide more extensive and higher-resolution data, enabling further refinement of theoretical models. Moreover, the development of new ultra-long wavelength observing facilities such as the DSL \citep{Shi2022a, Shi2022b, Chen2023}, LARAF \citep{Chen2024}, or FARSIDE \citep{Burns2021} is expected to bring in data at even lower frequencies, which would yield valuable insights, potentially leading to breakthroughs in understanding the physical mechanisms and evolutionary processes governing PS sources. The continued integration of observational data with advanced modeling efforts will be essential in resolving the outstanding questions surrounding these enigmatic sources.

\section*{Acknowledgments}
We thank the referees for constructive and detailed comments to improve the article.
We thank J. R. Callingham, Mengfan He, and Wenxiu Yang for their valuable discussions.
We acknowledge the support of the National SKA Program of China (2022SKA0110200, 2022SKA0110203), National Key R\&D Program of China grant no. 2022YFF0504300, 
and the National Natural Science Foundation of China (Nos. 12473091, 11975072, 11835009,12361141814). FXA acknowledges the support from the National Natural Science Foundation of China(12303016) and the Natural Science Foundation of Jiangsu Province (BK20242115).
We thank ChatGPT for assisting us in polishing the manuscript and optimizing the code to 
improve its execution speed.

\section*{Data Availability}

%

The data used in this work are all publicly available, the catalog presented in this article is summarized in Appendix \ref{s:appendixA} and is published online with this article as online-only material.
The dataset used in this study is available via\dataset[DOI: 10.5281/zenodo.14636864]{https://doi.org/10.5281/zenodo.14636864}.

\vspace{5mm}
\facilities{VLA(VLSSr, VCSS, NVSS, and VLASS), MWA(GLEAM), MOST(SUMSS), ASKAP(RACS)}


\software{astropy \citep{Astropy2013, Astropy2018, Astropy2022},  
          corner \citep{corner},
          emcee \citep{emcee2013},
          puma \citep[][]{puma2017}
          }



\appendix

\section{Catalog}\label{s:appendixA}
The column numbers, names, descriptions, and units in the tables represent different peaked-spectrum source samples \citep{Sun2025}. 
For sources with spectral peaks above \GHz, the peak shape is poorly fitted due to the lack of high-frequency data points. 
Although we have retained these data, we recommend using them with caution until more high-frequency observations become available. 
Additionally, we provide the calculated extent factor $ab/(a_{\rm psf}b_{\rm psf})$ to help exclude potential unresolved sources.
The complete data tables are available online via\dataset[DOI: 10.5281/zenodo.14636864]{https://doi.org/10.5281/zenodo.14636864}.

\begin{longtable*}{cllc}
	\caption{Catalog column descriptions}
	\label{tab:appendix}\\
		\hline
		Col. \# & Label & Description & Units\\
		\hline
        \endhead
        \hline
        \endfoot
        1 & GLEAM name & Name of the source in GLEAM & \\
        2 & GLEAM R.A. & R.A. of the source in GLEAM (J2000) & deg \\
        3 & GLEAM Decl. & Decl. of the GLEAM source in GLEAM (J2000) & deg \\
        4 & N\verb|_|SNR3 & Number of frequency measurements in the GLEAM survey with a signal-to-noise ratio greater than 3 & \\
        5 & S\verb|_|p & Flux density at the spectral peak & \Jy\\
        6 & e\verb|_|S\verb|_|p & Uncertainty in the spectral peak flux density& \Jy\\
        7 & nu\verb|_|p & Frequency of the spectral peak & \MHz\\
        8 & e\verb|_|nu\verb|_|p & Uncertainty in the spectral peak frequency & \MHz\\
        9 & alpha\verb|_|thin & Optically thin spectral index from ﬁtting Equation~(\ref{eq:pspmodel}) to the entire spectrum & \\
        10 & e\verb|_|alpha\verb|_|thin & Uncertainty in the optically thin spectral index & \\
        11& alpha\verb|_|thick & Optically thick spectral index predicted by the model fitting & \\
        12& e\verb|_|alpha\verb|_|thick & Uncertainty in the optically thick spectral index & \\
        13& K & normalization constant of flux density from ﬁtting Equation~(\ref{eq:pspmodel}) to the entire spectrum & \Jy\\
        14& e\verb|_|K & Uncertainty in the flux density normalization constant & \Jy\\
        15& nu\verb|_|ex & Characteristic frequency of external FFA from ﬁtting Equation~(\ref{eq:pspmodel}) to the entire spectrum & \MHz\\
        16& e\verb|_|nu\verb|_|ex& Uncertainty in the external FFA Characteristic frequency& \MHz\\
        17& nu\verb|_|SSA & Characteristic frequency of SSA from ﬁtting Equation~(\ref{eq:pspmodel}) to the entire spectrum & \MHz\\
        18& e\verb|_|nu\verb|_|SSA& Uncertainty in the SSA Characteristic frequency& \MHz\\
        19& nu\verb|_|FFA & Characteristic frequency of FFA from ﬁtting Equation~(\ref{eq:pspmodel}) to the entire spectrum & \MHz\\
        20& e\verb|_|nu\verb|_|FFA& Uncertainty in the FFA Characteristic frequency& \MHz\\
        21& p & p value from the F-test comparing the multi-mechanism hybrid model to the power-law model & \\
        22& BIC\verb|_|MMH & BIC value of the multi-mechanism hybrid model & \\
        23& BIC\verb|_|PL & BIC value of the power-law model & \\
        24& chi2 & chi2 value of the multi-mechanism hybrid model & \\
        25 & EF &  Extent factor $ab/(a_{\rm psf}b_{\rm psf})$ & \\
        26& S\verb|_|1$^{*}$  & Fux density at 1 \MHz predicted by the model fitting & \Jy \\
        27& e\verb|_|S\verb|_|1 & Uncertainty in the flux density at 1 \MHz & \Jy\\
        28& S\verb|_|10 & Fux density at 10 \MHz predicted by the model fitting & \Jy \\
        29& e\verb|_|S\verb|_|10 & Uncertainty in the flux density at 10 \MHz & \Jy\\
        30& S\verb|_|30 & Fux density at 30 \MHz predicted by the model fitting & \Jy \\
        31& e\verb|_|S\verb|_|30 & Uncertainty in the flux density at 30 \MHz & \Jy\\
        32& VLSSr R.A. & R.A. of the source in VLSSr (J2000) & deg \\
        33& VLSSr Decl. & Decl. of the source in VLSSr (J2000) & deg \\
        34&S\verb|_|VLSSr & VLSSr ﬂux density at 74 \MHz & \Jy\\
        35&e\verb|_|S\verb|_|VLSSr & Uncertainty in the VLSSr ﬂux density & \Jy\\
        36& TGSS-ADR1 name & name of the source in TGSS-ADR1 & \\
        37& TGSS R.A. & R.A. of the source in TGSS-ADR1 (J2000) & deg \\
        38& TGSS Decl. & Decl. of the source in TGSS-ADR1 (J2000) & deg \\
        39&S\verb|_|TGSS & TGSS ﬂux density at 150 \MHz & \Jy\\
        40&e\verb|_|S\verb|_|TGSS & Uncertainty in the TGSS ﬂux density & \Jy\\
        41&VCSS name & Name of the source in VCSS & \\
        42& VCSS R.A. & R.A. of the source in VCSS (J2000) & deg \\
        43& VCSS Decl. & Decl. of the source in VCSS (J2000) & deg \\
        44&S\verb|_|VCSS & VCSS ﬂux density at 340 \MHz & \Jy\\
        45&e\verb|_|S\verb|_|VCSS & Uncertainty in the MRC ﬂux density & \Jy\\
        46&MRC name & Name of the source in MRC & \\
        47& MRC R.A. & R.A. of the source in MRC (J2000) & deg \\
        48& MRC Decl. & Decl. of the source in MRC (J2000) & deg \\
        49&S\verb|_|MRC & MRC ﬂux density at 408 \MHz & \Jy\\
        50&e\verb|_|S\verb|_|MRC & Uncertainty in the MRC ﬂux density & \Jy\\
        51& SUMSS R.A. & R.A. of the source in SUMSS (J2000) & deg \\
        52& SUMSS Decl. & Decl. of the source in SUMSS (J2000) & deg \\
        53&S\verb|_|SUMSS & SUMSS ﬂux density at 843 \MHz & \Jy\\
        54&e\verb|_|S\verb|_|SUMSS & Uncertainty in the SUMSS ﬂux density & \Jy\\
        55&RACS name & Name of the source in RACS & \\
        56& RACS R.A. & R.A. of the source in RACS (J2000) & deg \\
        57& RACS Decl. & Decl. of the source in RACS (J2000) & deg \\
        58&S\verb|_|RACS & RACS ﬂux density at 887.5 \MHz & \Jy\\
        59&e\verb|_|S\verb|_|RACS & Uncertainty in the RACS ﬂux density$^{**}$ & \Jy\\
        60&NVSS name & Name of the source in NVSS & \\
        61& NVSS R.A. & R.A. of the source in NVSS (J2000) & deg \\
        62& NVSS Decl. & Decl. of the source in NVSS (J2000) & deg \\
        63&S\verb|_|NVSS & NVSS ﬂux density at 1400 \MHz & \Jy\\
        64&e\verb|_|S\verb|_|NVSS & Uncertainty in the NVSS ﬂux density & \Jy\\
        65&VLASS name & Name of the source in VLASS & \\
        66& VLASS R.A. & R.A. of the source in VLASS (J2000) & deg \\
        67& VLASS Decl. & Decl. of the source in VLASS (J2000) & deg \\
        68&S\verb|_|VLASS & VLASS ﬂux density at 3000 \MHz & \Jy\\
        69&e\verb|_|S\verb|_|VLASS & Uncertainty in the VLASS ﬂux density & \Jy\\
        70& $z$ & Spectroscopic redshift in \cite{Gordon2023} & \\
        71& e\verb|_|$z$ & Uncertainty in the Spectroscopic redshift & \\
        72& S\verb|_|76 & GLEAM flux density at 76 \MHz & \Jy\\
        73& e\verb|_|S\verb|_|76 & Uncertainty in the flux density at 76 \MHz & \Jy\\
        74& S\verb|_|84 & GLEAM flux density at 84 \MHz & \Jy\\
        75& e\verb|_|S\verb|_|84 & Uncertainty in the flux density at 84 \MHz & \Jy\\
        76& S\verb|_|92 & GLEAM flux density at 92 \MHz & \Jy\\
        77& e\verb|_|S\verb|_|92 & Uncertainty in the flux density at 92 \MHz & \Jy\\
        78& S\verb|_|99 & GLEAM flux density at 99 \MHz & \Jy\\
        79& e\verb|_|S\verb|_|99 & Uncertainty in the flux density at 99 \MHz & \Jy\\
        80& S\verb|_|107 & GLEAM flux density at 107 \MHz & \Jy\\
        81& e\verb|_|S\verb|_|107 & Uncertainty in the flux density at 107 \MHz & \Jy\\
        82& S\verb|_|115 & GLEAM flux density at 115 \MHz & \Jy\\
        83& e\verb|_|S\verb|_|115 & Uncertainty in the flux density at 115 \MHz & \Jy\\
        84& S\verb|_|122 & GLEAM flux density at 122 \MHz & \Jy\\
        85& e\verb|_|S\verb|_|122 & Uncertainty in the flux density at 122 \MHz & \Jy\\
        86& S\verb|_|130 & GLEAM flux density at 130 \MHz & \Jy\\
        87& e\verb|_|S\verb|_|130 & Uncertainty in the flux density at 130 \MHz & \Jy\\
        88& S\verb|_|143 & GLEAM flux density at 143 \MHz & \Jy\\
        89& e\verb|_|S\verb|_|143 & Uncertainty in the flux density at 143 \MHz & \Jy\\
        90& S\verb|_|151 & GLEAM flux density at 151 \MHz & \Jy\\
        91& e\verb|_|S\verb|_|151 & Uncertainty in the flux density at 151 \MHz & \Jy\\
        92& S\verb|_|158 & GLEAM flux density at 158 \MHz & \Jy\\
        93& e\verb|_|S\verb|_|158 & Uncertainty in the flux density at 158 \MHz & \Jy\\
        94& S\verb|_|166 & GLEAM flux density at 166 \MHz & \Jy\\
        95& e\verb|_|S\verb|_|166 & Uncertainty in the flux density at 166 \MHz & \Jy\\
        96& S\verb|_|174 & GLEAM flux density at 174 \MHz & \Jy\\
        97& e\verb|_|S\verb|_|174 & Uncertainty in the flux density at 174 \MHz & \Jy\\
        98& S\verb|_|181 & GLEAM flux density at 181 \MHz & \Jy\\
        99& e\verb|_|S\verb|_|181 & Uncertainty in the flux density at 181 \MHz & \Jy\\
        100& S\verb|_|189 & GLEAM flux density at 189 \MHz & \Jy\\
        101& e\verb|_|S\verb|_|189 & Uncertainty in the flux density at 189 \MHz & \Jy\\
        102& S\verb|_|197 & GLEAM flux density at 197 \MHz & \Jy\\
        103& e\verb|_|S\verb|_|197 & Uncertainty in the flux density at 197 \MHz & \Jy\\
        104& S\verb|_|204 & GLEAM flux density at 204 \MHz & \Jy\\
        105& e\verb|_|S\verb|_|204 & Uncertainty in the flux density at 204 \MHz & \Jy\\
        106& S\verb|_|212 & GLEAM flux density at 212 \MHz & \Jy\\
        107& e\verb|_|S\verb|_|212 & Uncertainty in the flux density at 212 \MHz & \Jy\\
        108& S\verb|_|220 & GLEAM flux density at 220 \MHz & \Jy\\
        109& e\verb|_|S\verb|_|220 & Uncertainty in the flux density at 220 \MHz & \Jy\\
        110& S\verb|_|227 & GLEAM flux density at 227 \MHz & \Jy\\
        111& e\verb|_|S\verb|_|227 & Uncertainty in the flux density at 227 \MHz & \Jy\\
        112& S\verb|_|200,wide & Flux density of the source in the GLEAM wideband flux density  & \Jy\\
        113& e\verb|_|S\verb|_|200,wide & Uncertainty in the GLEAM wideband flux density & \Jy\\
        \hline
        \multicolumn{4}{l}{$^{*}$ For some sources, the flux density is below the detection limit of the instrument and recorded as 0.}\\
        \multicolumn{4}{l}{$^{**}$ The ``E\_Total\_flux\_Source" error in RACS, which is the combined error on the total flux density.}\\
        \multicolumn{4}{l}{The full catalog is available in the online journal and via\dataset[DOI: 10.5281/zenodo.14636864]{https://doi.org/10.5281/zenodo.14636864}.}\\
\end{longtable*}


\bibliography{main}{}
\bibliographystyle{aasjournal}



\end{document}